\documentclass[a4paper,12pt]{article}
\usepackage{graphicx} 
\usepackage{epstopdf}
\usepackage{subfigure}

\begin{document}

\hoffset = -1truecm \voffset = -2truecm \baselineskip = 10 mm

\title{Looking for the possible gluon condensation signature in sub-TeV gamma-ray spectra: from active galactic nuclei
to gamma ray bursts}

\author{Wei Zhu$^a$, Zechun Zheng$^a$, Peng Liu$^a$, Lihong Wan$^a$,\\
          Jianhong Ruan$^a$ and Fan Wang$^b$
        \\
        \normalsize $^a$Department of Physics, East China Normal University,
        Shanghai 200241, China \\
        \normalsize $^b$Department of Physics, Nanjing University,
        Nanjing,210093, China\\
    }

\date{}

\newpage

\maketitle

\vskip 3truecm

\begin{abstract}
The gluon condensation in the proton as a dynamical model is used to
treat a series of unsolved puzzles in sub-TeV gamma ray spectra,
they include the broken power-law of blazar's radiation, the
hardening confusion of 1ES 1426+428, Mkn 501, and the recently
recorded sub-TeV gamma spectra of GRB 180720B and GRB 190114C. We
find that the above anomalous phenomena in gamma ray energy spectra
can be understood with the simple broken power law based on a QCD
gluon condensation effect.

\end{abstract}

    {\bf keywords}:cosmic ray theory, gamma ray theory, very high energy cosmic rays

        \vskip 1truecm

\newpage

\vskip 1truecm

\newpage
\begin{center}
\section{\bf Introduction}
\end{center}

 Gluons are Boson. A Quantum Chromodynamics (QCD) analysis shows that
the gluon distribution in the proton may evolute to a chaos solution
at high energy limit, which arouses the strong shadowing and
antishadowing effects and squeezes the gluons into a narrow space
near a critical momentum [1-3]. This is the gluon condensation (GC).

    The GC would lead to intriguing signatures in proton collisions,
provided the GC-threshold $E_p^{GC}$ enters the observable energy
region. However, we have not directly observed the GC-effect at the
Large Hadron Collider (LHC). Therefore, we turn to the astrophysical
and cosmological observations. The energy of the protons accelerated
in universe may exceed $E_p^{GC}$ and causes the GC-effect in the
collisions.

    In a previous work [4] we have used the GC model to explain the sharp broken power
 law of the $\gamma$-ray spectra
in supernova remnant (SNR) Tycho. In this work we will explore more
GC-examples in the $\gamma$-ray spectra of active galactic nuclei
(AGN).  AGN represents a large population of extragalactic objects
characterized with extremely luminous electromagnetic radiation
produced in very compact volumes. AGNs with relativistic directional
jets (so-called blazars) are very effective TeV $\gamma$-sources.
Therefore, blazars may provide an ideal laboratory for studying the
GC effect since they have extreme physical conditions.

    AGNs have relatively large redshift. TeV radiation from these
sources is affected by intergalactic absorption. Comparing the
observed AGN spectrum with their intrinsic (source) spectrum, one
can obtain an important information about the extragalactic
background light (EBL) at the infrared energy band. Extracting the
EBL structure from intrinsic and observed spectra, or alternatively,
using a reasonable EBL model to excavate the emission mechanism in
AGN are the hot subjects in astronomy.

      Very high energy (VHE) $\gamma$-ray spectra have the following features. (i) The spectral energy
distribution (SED) is peaked at GeV-TeV band; (ii) Radiations have
strong power, thus, $\gamma$-rays can reach the earth trough the EBL
absorption. (iii) The intrinsic spectra present a sharp broken
power-law after deducting the EBL corrections from the observed
spectra in a series AGN sources , which does not have any existing
radiation theory to explain. (iv) Some of them have an extra hard
tail at the TeV range, it even raises doubts about the Lorentz
covariance. (v) Many AGN examples present the above phenomenons and
it might imply a new general dynamics, which have never been
recognized before.

    A topic related to VHE gamma ray radiation is gamma ray burst (GRB).
GRBs are extremely violent, serendipitous sources of electromagnetic
radiation in the Universe. Recently, sub-TeV gamma-rays were
detected from GRB 190114C by MAGIC[5] and GRB 180720B by HESS
telescopes [6], respectively. This discovery caused great interest.
For this sake, we will discuss it using a same GC-framework in a
special section 5. We find that the SED of both these two events
have a series of characteristics of the GC-signature. We also show
that the VHE gamma spectra in GRB and AGN relate to a same GC-effect
but in two different cosmic environments.

      Organization of the article is as follows. We will
briefly review the hadronic mechanism for $\gamma$-ray spectra in
section 2, i.e., $p+p\rightarrow \pi$ and $\pi^0\rightarrow
2\gamma$, and show how the GC-process works in the hadronic
mechanism. After a straightforward, but general derivation, we
naturally give a sharp broken power-law in $\gamma$-ray spectra.
Then a simplified calculation method for the EBL is given in section
3. We chose ten examples of $\gamma$-ray spectra of blazars to
indicate the GC-effect in section 4. More complex hardening VHE
gamma-ray spectra are explained by using the same GC-model in
section 5. We discuss VHE spectra of GRB 180720B and GRB 190114C in
section 6. A summary is presented in last section.

\newpage
\begin{center}
\section{The GC-effect in high energy $\gamma$-ray spectra}
\end{center}

 The SED of
blazars is assumed to be dominated by the emission from a
relativistic jet pointing close to our line of sight.  The
characteristic SED of blazars shows two broad non-thermal well
defined continuum peaks. The first hump located between the infrared
(IR) and $X$-ray bands, whereas the second hump exhibits a maximum
at the $\gamma-$energy band. The origin of low energy peak is
attributed to the synchrotron emission of relativistic electrons in
the magnetic field of the jet. The productions of VHE $\gamma-$rays
have different explanations. In leptonic model high energy electrons
scatter on low energy photons through inverse Compton (IC)
scattering $e+\gamma_{low~energy} \rightarrow
e_{low~energy}+\gamma_{VHE}$ and form TeV-$\gamma$. Normally, these
low energy photons are produced in the environment of stars due to
thermal emission or due to synchrotron emission by the high energy
electrons in the ambient magnetic fields (the SSC model) [e.g. 7-9].
According to hadronic model, VHE $\gamma-$ray emissions are
dominated by neutral pion decay into photons in the following
cascade processes: $p+nuclus\rightarrow p'+\pi^0+others$ and
following $\pi^0\rightarrow 2\gamma$ [e.g. 10-12].

    Using hadronic mechanism the $\gamma$-flux $\Phi_{\gamma}$ can be
described as

$$\Phi_{\gamma}(E_{\gamma})=C_{\gamma}\left(\frac{E_{\gamma}}{E_0}\right)^{-\beta_{\gamma}}\int_{E_{\pi}^{min}}^{E_{\pi}^{max}}dE_{\pi}$$
$$\times\left(\frac{E_p}{E_p^{GC}}\right)^{-\beta_p}
{N_{\pi}(E_p,E_{\pi})
\frac{d\omega_{\pi-\gamma}(E_{\pi},E_{\gamma})}{dE_{\gamma}}},
\eqno(2.1)$$ where indexes $\beta_{\gamma}$ and $\beta_p$ denote the
propagating loss of gamma rays inside the source and the
acceleration process of protons respectively; $C_{\gamma}$
incorporates the kinematic factor with the flux dimension and the
percentage of $\pi^0\rightarrow 2\gamma$.

       The spectrum of gamma-rays from $\pi^0$ decay
in the center of mass system will have the maximum
 at $E_{\gamma}=m_{\pi}c^2/2$, independent of
 the energy distribution of $\pi^0$ mesons and consequently of the parent protons. Usually,
 the distribution of $N_{\pi}$
 is parameterized by using the $pp$ cross-section $\sigma_{pp}$, the resulting distribution $\Phi_{\gamma}$
 presents a smooth excess near $E_{\gamma}\sim 1GeV$. This prediction has been
 proven
 by the $\gamma$-ray spectra of supernova remnants (SNRs) [13] (see figure 1).
 In a broad GeV-TeV region the total cross-section $\sigma_{pp}\sim 30~mb$, while the
 angle-averaged total cross-section of inverse Compton scattering
 $\sigma_{IC}\sim (8\pi/3)r_e^2\sim 800~mb$, $r_e\sim 3\times
 10^{-13}~cm$. Therefore, the lepton mechanism has a stronger radiation power than the hadronic
 mechanism.

     However, the GC-effect produces a different spectral structure
in the same hadronic model. According to QCD, the number of
secondary particles (they are mostly pions) at the high energy $pp$
collisions relates to how much gluons participate into the
multi-interactions. Pions will rapidly grow from $E_{\pi}^{GC}$
since a lot of condensed gluons enter the interaction range. One can
image that the number of pion in this case reaches its maximum
value, i.e., all available kinetic energy of the collision at the
center-of-mass system is almost used to create pions (see figure 2).
Using general relativistic invariant and energy conservation, we
straightforwardly obtain the solution $N_{\pi}$ in $pp$ collisions
using $GeV$-unit [4, 14].

$$\ln N_{\pi}=0.5\ln E_p+a, ~~\ln N_{\pi}=\ln E_{\pi}+b,  \eqno(2.2)$$
$$~~ where~E_{\pi}
\in [E_{\pi}^{GC},E_{\pi}^{max}].$$ The parameters are

$$a\equiv 0.5\ln (2m_p)-\ln m_{\pi}+\ln K, \eqno(2.3)$$ and

$$b\equiv \ln (2m_p)-2\ln m_{\pi}+\ln K, \eqno(2.4)$$ $K$ is inelasticity.
Equation (2.2) gives the relations among $N_{\pi}$, $E_p$ and
$E_{\pi}^{GC}$ by one-to-one, it leads to the following
GC-characteristic spectrum.

    Substituting them into equation (2.1), we have [4,14]

$$E_{\gamma}^2\Phi^{GC}_{\gamma}(E_{\gamma})$$
$$=C_{\gamma}\left(\frac{E_{\gamma}}{E_{\pi}^{GC}}\right)^{-\beta_{\gamma}}
\int_{E_{\pi}^{GC}~or~E_{\gamma}}^{E_{\pi}^{GC,max}}dE_{\pi}$$
$$\times{\left(\frac{E_p}{E_p^{GC}}\right)^{-\beta_p}N_{\pi}(E_p,E_{\pi})
\frac{2}{\beta_{\pi}E_{\pi}}}$$
$$=\left\{
\begin{array}{ll}
\frac{2C_{\gamma}}{2\beta_p-1}e^b(E_{\pi}^{GC})^3\left(\frac{E_{\gamma}}{E_{\pi}^{GC}}\right)^{-\beta_{\gamma}+2} & {\rm if~}E_{\gamma}\leq E_{\pi}^{GC}\\\\
\frac{2C_{\gamma}}{2\beta_p-1}e^b(E_{\pi}^{GC})^3\left(\frac{E_{\gamma}}{E_{\pi}^{GC}}\right)^{-\beta_{\gamma}-2\beta_p+3}
& {\rm if~} E_{\gamma}>E_{\pi}^{GC}
\end{array} \right. $$
$$\equiv\left\{
\begin{array}{ll}
\Phi_0\left(\frac{E_{\gamma}}{E_{\pi}^{GC}}\right)^{-\Gamma_1+2} & {\rm if~}E_{\gamma}\leq E_{\pi}^{GC}\\\\
\Phi_0\left(\frac{E_{\gamma}}{E_{\pi}^{GC}}\right)^{-\Gamma_2+2} &
{\rm if~} E_{\gamma}>E_{\pi}^{GC}
\end{array} \right. ,\eqno(2.5)$$where the lower limit of the integration takes
$E_{\pi}^{GC}$ (or $E_{\gamma}$) if $E_{\gamma}\leq E_{\pi}^{GC}$
(or if $E_{\gamma}> E_{\pi}^{GC}$). Equation (2.5) is a typical
broken power-law, it is formed by the GC-effect in the pion
distribution $N_{\pi}$ rather than the decay mechanism
$\pi^0\rightarrow 2\gamma$. This solution is different from all
other well known radiation spectra. We regard it as the GC-character
(see figure 2). Note that comparing with the hadronic model without
the GC effect, the total cross section $\sigma_{pp}$ has increased
by several orders of magnitude.

     The value of $\beta_{\gamma}$ in equation (2.5) may approximately take
the value of $0\sim 2$, where zero means that the energy loss of
$\gamma$-ray is almost negligible when it travel inside the source.
The term $E_p^{-\beta_p}$ is from the contributions of the
integrated initial proton flux $\int dE_pJ_p$, where $J_p\sim
E_p^{-\Gamma_p}$ and $\beta_p=\Gamma_p-1$. Usually the index
$\Gamma_p\sim 3$ in the observation on Earth. On the other hand, a
more harder proton beam inside the source is possible but equation
(2.5) requests that $\beta_p>0.5$. Therefore, we take $\beta_p\in
[0.5,2]$. Thus, equation (2.5) predicts that the change of spectra
index on two sides of the break takes $\Gamma_2-\Gamma_1
=2\beta_p-1\simeq 0\sim 3$, which is measurable and irrelevant to
the value of $\beta_{\gamma}$.

    The GC-threshold $E_{\pi}^{GC}$ is nuclear number $A$-dependent: $E_{\pi}^{GC}(p-p)>E_{\pi}^{GC}(A-A)$
since the nonlinear corrections enhance with increasing $A$. A rough
estimation finds that the values of $E_{\pi}^{GC}$ may distribute in
a range $0.1TeV\sim 30TeV$ for the collisions between heavy nuclei
and $pp$ collision [e.g. 15]. Generally, we will meet different
values of $E_{\pi}^{GC}$ for different $\gamma$-ray sources, the
result relates to the dominate component of the hadron beam. While
we lack a reliable theoretical prediction about them. We expect more
information from the AGN observations, which may help us to
establish a rule about $E_{\pi}^{GC}\sim A$.

    The above discussion is focused on a single GC-source.
The observed spectra in sky survey may origin from several
GC-sources in AGN source. We will discuss them in section 5.

\newpage
\begin{center}
\section{The EBL corrections to the gamma ray spectra}
\end{center}

The EBL is the cosmic background photon field, which is mainly
produced by stars and interstellar medium in galaxies throughout the
cosmic history. The EBL could be directly measured with different
instruments. However, the foreground zodiacal light and galactic
light may introduce large uncertainties in such measurements and
make it difficult to isolate the EBL contribution from the observed
multi-TeV flux from distant blazars. In this sense, the
determination of the intrinsic $\gamma$-ray spectra helps to find a
correct EBL model.

    We list the EBL formula, which modify the propagation of VHE $\gamma$-rays traveling through
intergalactic space from the sources. In process
$\gamma_{VHE}+\gamma_{EBL}\rightarrow e^-+e^+$, the observed
$\gamma$-ray flux on Earth is related to the intrinsic flux of the
source as refs. [16, 17].

$$\Phi_{\gamma}^{ob}(E_{\gamma})=\Phi_{\gamma}^{in}(E_{\gamma})\times e^{-\tau}. \eqno(3.1)$$
The optical depth $\tau$ reads

$$\tau=\int_0^R\int_{\lambda_{min}}^{\lambda_{max}}\sigma_{\gamma\gamma}(E_{\gamma},\lambda)\lambda F_{\lambda}(
\lambda, r)d\lambda dr,\eqno(3.2)$$where $\lambda
F_{\lambda}(\lambda, r)$ describes the spectral and spatial
distribution of the target photon field in sky.

    Because of the narrowness of the cross section
$\sigma_{\gamma\gamma}$, equation (3.2) can be simplified as

$$\tau\simeq A\left(\frac{\lambda^* F_{\lambda^*}}{1nW/m^2sr}\right)\left(\frac{E_{\gamma}}{1TeV}\right)\left
(\frac{z}{z_0}\right), \eqno(3.3)$$where
$\lambda^*=1.4(E_{\gamma}/1TeV)\mu m$ and the Hubble constant has
been included into the coefficient $A$. Usually, $A$ is regarded as
a constant. However, we find that such a rough estimation cannot
extract the correct intrinsic flux. As an improvement, we take $A$
as a function of $E_{\gamma}$. In next section we will use the data
of PKS 2155-304 blazar ($z_0=0.116$) to obtain a relation of $A\sim
E_{\gamma}$ (see figure 4). Assuming this relation is fixed for the
sources with not too larger value $z<1$, one can get a reasonable
relation between $\Phi_{\gamma}^{in}$ and $\tau$.

    Usually the distributions of the EBL at near infrared have different
categories. We take a distribution $\lambda^* F_{\lambda^*}$ (see
figure 3), which is proposed by [18] and it is roughly the low bound
of the EBL by the estimation based on the deep-galaxy-surveys data.
We call it as the lower EBL model.

\newpage
\begin{center}
\section{Gamma ray spectra with single GC-source
}
\end{center}

We use equation (2.5) and the lower EBL model in figure 3 to extract
$\Psi_{\gamma}^{in}$ from the observed data. Note that a small
modification in $\Phi_{\gamma}^{ob}$ at $E_{\gamma}\gg 0.1TeV$ may
arise a large deformation in $\Phi_{\gamma}^{in}$ since the
amplification of the factor $exp(-\tau)$ in equation (3.1).
Therefore, the actual approach is using a suitable distribution
$\Phi^{GC}_{\gamma}$ in equation (2.5) to fit the observed data (the
dashed curves) in figure 5. Note that equation (2.5) has only four
free parameters, which are much smaller than other radiation models.

    Figure 5 shows the SEDs of ten blazars in the GC-model (see solid lines).
They all present the sharp broken power-law. The relating parameters
are listed in table 1, where $\Gamma_1\equiv \beta_{\gamma}$ and
$\Gamma_2\equiv \beta_{\gamma}+2\beta_p-1$. One can see that
although the EBL may change the index of the spectrum at
$E_{\gamma}>E_{\pi}^{GC}$, $\Phi_{\gamma}^{in}$ always shows a sharp
broken power-law. The different EBL may deform $\Phi_{\gamma}^{in}$,
however, Korochkin, Neronov and Semikoz [21] found that the
intrinsic spectra are still sharply broken after the corrections of
the various EBL models. Therefore, we suppose that the GC is the
dynamics of the broken power-law.

     The GC-effect not only exists in blazars, but also appears in non-blazar
AGNs. PKS 0625+354 has not a clear evidence for optical blazar
characteristics [27], however its intrinsic $\gamma$-ray spectrum in
figure 5 still presents a broken power-law.

    Blazars are known for their variability on a wide range of
timescales. Most studies of TeV gamma-ray blazars focus on short
timescales.  The observations of the blazar 1ES 1215+303 from
2008-2017 are investigated by a combining Fermi-LAT and VERITAS
collaborations [28]. The results show that the parameters of the
observed spectra change in a narrow range, and a sharp broken
power-law is still kept.

     The intrinsic $\gamma$-ray spectra will have different forms if
different EBL models are used for an observed spectrum. PKS 2155-304
and 1ES 1218+304 are located at redshifts $z=0.116$ and $0.1824$,
respectively. Some authors expect that they have the unusually hard
and unbreakable intrinsic spectra due to a strong EBL absorption
[25,29]. We take 1ES 0229+200 as an example. A stronger EBL model
with an additional narrow Gaussian component (dashed curve in figure
6) may fit the data (see figure 7). However, similar TeV spectra of
PKS 0625+354, 1ES 1959+650 and 1ES 1218+304 in figure 5 will deviate
from the theoretical predictions if the same stronger EBL model is
used. In other words, different EBL distributions are needed if all
broken power behaviors in AGNs are attributed to the EBL absorption.
It means that the EBL would loss its universality. On the other
hand, the lower EBL model uniformly describes all blazar spectra in
figure 5. Thus, from the perspective of the universality of the EBL,
we adopt the lower EBL model.

\newpage
\begin{center}
\section{Hardening VHE gamma-ray spectra
}
\end{center}

As well known that the index of a cosmic ray spectrum at the
asymptotic range decreases with increasing energy. While some of
TeV-blazars present an opposite trend in their intrinsic spectra,
i.e., the spectra have hard tails. To understand this interesting
phenomenon, the photon to axion-like particle (ALP) conversion, and
even the Lorentz invariance violation are proposed.

    We have mentioned at section 2, the GC-threshold in
the $pp$ collision is much bigger than that in the $p-$heavy nuclei
collisions since the GC-critical value $x_c$ is nuclear mass
$A$-dependent. We can observe multi-values of $E_{\pi}^{GC}$ if a
gamma ray source contains $pp$- and $pA$- (or $AA$-) collisions.
Usually we only record a single GC-source, because the observation
energy range is not wide enough, or some of signals are too weak. We
will show that the double-GC effect may explain the hardening tail.
We give a few of hardening examples in figure 8 and table 2.

    1ES 1426+428 is a TeV $\gamma$-ray source that has been noticed
earlier. The TeV $\gamma$-rays from this source arrive the earth
after the EBL absorption. However, the measured spectrum looks
significantly different compared to the other blazar spectra since
it has a bump at $E_{\gamma}\sim 1~TeV$ [36]. We use the double
GC-sources to explain 1ES 1426+428 spectrum. The parameters
$\beta_{\gamma}^{II}=0$ is much smaller than that
$\beta_{\gamma}^{I}=0.96$ and $E_{\pi}^{GCII}=10~TeV\gg
E_{\pi}^{GCI}=0.3~TeV$. This result can be understood as follows:
the first GC-source produced in the heavy nuclear collisions deep in
the star with a smaller $E_{\pi}^{GC}$, the spectrum has non-zero
value of  $\beta_{\gamma}^I$, while the second GC-source produced by
the light nuclear collisions with a larger $E_{\pi}^{GC}$ at the
surface of this star, where the photons go directly into the sky
without passing through the polluted range, therefore, its spectrum
has $\beta_{\gamma}^{II}\sim 0$. One can find that superposition of
the first intrinsic spectrum at $E_{\gamma}>E_{\pi}^{GCI}$ (where
the index $\Gamma_2^I-2>0$) and the second intrinsic spectrum at
$E_{\gamma}<E_{\pi}^{GCII}$ (where the index $\Gamma_1^{II}-2<0$)
causes a hardening tail (see red dashed curve in figure 8) even
considering the EBL correction.

    TeV blazar Markarian (Mkn) 501 is a striking star. Although its spectrum can be described, at least qualitatively,
by the leptonic and normal hadronic models, it is difficult to
explain a pile-up at the end of the absorption-corrected spectrum.
For this sake, several extreme hypotheses have been proposed to
overcome this so called "IR-background-TeV gamma ray crisis". We try
to answer this question. We consider that the leptonic mechanism
dominates the spectrum of Mkn 501 at $0.1TeV-10TeV$, and it can be
described by a curvilinear form $\sim 1.26 \times 10^{-10}
E^{-0.01}_{\gamma}exp(-E_{\gamma}/30TeV)$. However, a GC-effect
participated in the process at $E_{\gamma}>10TeV$. The result is
shown in figure 8, where we assume temporarily the parameters
$\beta_p\simeq 1$ and $E_{\pi}^{GC}\simeq 30~TeV$ because we lack
the data at higher energy.

\newpage
\begin{center}
\section{GRB 180720B and GRB 190114C}
\end{center}

    GRBs occur suddenly and unpredictably with a rate of approximately
one burst per day. The durations of the GRBs are very short ranging
from fractions of a second up to 100 seconds. The GRBs appear to be
uniformly distributed over the whole sky. The short burst duration
it is very difficult to identify a GRB event with a known object.

    The generally accepted the sub-TeV emission mechanism of the GRB is the SSC
mechanism synchrotron, where a synchrotron-emitting source must
produce high energy radiation through up-scattering of synchrotron
photons by the same electrons [37-39]. Thus, sub-TeV radiation is
expected from GRB afterglows at early stage. However, several
attempts to detect very high energy (VHE) ($>100~GeV$) gamma-rays
from GRBs were unsuccessful, resulting only in upper limits. Just
recently, VHE photons were detected in GRB 180720B and GRB 190114C.

    The connection between the VHE emission by IC and the
low energy afterglows in synchrotron radiation does not seem
inevitable, since most of the GRB events have not the VHE spectral
record. we do not exclude the possibility existing a new radiation
mechanism in GRB. We try to use the GC-model to understand the VHE
spectra from these two GRBs.

    Considering a proton- (or nuclei-) beam after being accelerated to
an extremely high energy in a AGN and traveling in interstellar
space. If it randomly collides with high-density matter (meteorite,
small star even mini black hole), electromagnetic radiation will be
generated in the short time of collision, which includes VHE
gamma-rays if the collision energy beyond the GC-threshold. These
are called as the short-duration bursts

    Using equation (3.1) we extract $\Phi_{\gamma}^{in}$ for GRB 190114C
(solid curve in figure 9). According to equation (2.5) we have the
parameters $\beta_{\gamma}=0$, $\beta_p=2.025$ and
$E_{\pi}^{GC}=0.2~TeV$, where we consider that the energy loss of
gamma-rays inside small target is negligible (i.e.,
$\beta_{\gamma}\sim 0$). Note that no signal from GRB 190114C was
detected when $E_{\gamma}<0.2~TeV$. We use the result in ref. [40]
to place the upper limits on its SED (see horizontal bars with
downwards arrows), where the instrument resolution and a special
program are considered. The result of the GC-model is acceptable.

    A similar VHE spectrum of RGB 180720B is given in figure 10.
Note that the values $\beta_p$ for RGB 180720B and GRB 190114C are
obviously larger than the maximum value of $\beta_p\sim 1.5$ in the
AGNs (see tables 1 and 2). It confirms the following fact: $\beta_p$
in the GRB-events includes the corrections due to the energy loss of
the proton beam in the interstellar journey, therefore, $\beta_p$
has a larger value. Note that MAGIC used a higher EBL model to get a
hard spectrum $\Phi_{\gamma}^{in}\sim E_{\gamma}^{-2.22}$ for GRB
190114C [5], which corresponds to $\beta_p=1.61$.

    One can find that the result of the GC model is different from the
predictions of the SSC model. The former shows a typical broken
power-law in a broad energy range without the cut factor. On the
other hand, both the SSC and the traditional hadronic model fit the
gamma ray spectra using a log-parabolic curve
$E_{\gamma}^{-\Gamma-a\log(E_{\gamma}/E_{cut})}$ with a cut factor.

    The suddenly increasing $pp$ cross section leads to a big
excess of $\gamma$ number at the GC-threshold $E_{\pi}^{GC}$. This
greatly improves the radiative efficiency in the conversion from
kinetic energy of the accelerated protons to the radiation energy.
Besides, this is exactly what the GRB-event needs.

   If the VHE gamma-ray spectra of both AGN (blazar) and GRB are dominated by the
GC-effect, they have a similar broken power-law. However,
acceleration and collisions of the protons in the blazar occur in a
same AGN-source, the observed spectra can maintain a long time,
although its intensity changes over time. On the other hand, GRB may
be a product of extremely fast protons hitting matter during
interstellar travel. The recorded GRB spectra are random in a very
short formation time. These are well known but have no satisfactory
explanation yet.

\newpage
\begin{center}
\section{Summary}
\end{center}

 The hadronic model and leptonic model are two major categories
of theories explaining the VHE $\gamma$-ray spectra. The GC model is
a hadronic model considering the GC-effect. As we have proven that
the GC origins from a general QCD evolution dynamics, once a
$\gamma$-ray spectrum with the GC characteristic is confirmed, a
series of similar GC spectra are found to be inevitable. We
collected some of typical spectra in the large amount of existing
references. For example, the spectra in figures 5 and 8 are
presented as the examples in a summary article [24] and we adopted
them all. We find that the change of these spectra indexes before
and after breaking is within the theoretical value range of the GC
model. This is a strong support for the GC model and it may be the
origin of the broken power law. Of cause, this does not exclude
other hadronic and leptonic models. As we know that the broken power
law is often used to fit the VHE $\gamma$-ray spectra as a
mathematical parameter formula. The GC model gives this formula a
deep physical significance and predictive power.

    The gamma ray spectra with GC-effect has a typical broken power
form, which is clearly different from the traditional hadronic model
and leptonic model. Note that although the distribution
$E^2_{\gamma}\Phi_{\gamma}$ turns around the break from an increase
to a downward, however, the SED of $\Phi_{\gamma}$ almost decrease
with increasing energy since both $\Gamma_1\geq 0$ and $\Gamma_2>0$.
Besides, the slopes of the $\gamma$ spectrum on two sides of the
break are correlated by $\Gamma_2-\Gamma_1=2\beta_p-1\in [0, 3]$ as
shown in tables 1 and 2. One can find that the VHE $\gamma$-ray
spectra have different break forms in a reasonable range of the
parameters $\beta_{\gamma}$ and $\beta_p$ (see figure 11).
Especially, they may combine a more complex spectra in figure 8.

    The application condition of the the GC model is that the initial
protons have enough high energy, thus, the GC-threshold $E_p^{GC}$
may enters the observable energy region at $pp$ collision. For
example, the observation of a $\gamma$ signal at
$E_{\gamma}=100~GeV$ needs the incident energy about $E_p\sim
10~TeV$ in the traditional hadronic model [17]. However, the GC
effect increases the number of the pions, but it is at the cost of
consuming more proton's kinetic energy. Using equation (2.2), the
protons should be accelerated beyond $E_p=E_{\pi}^2\exp[2(b-a)]\sim
10^{15}eV$ for producing gamma signal at $E_{\gamma}\sim 100~GeV$.
Moreover, this value is only a lower limit and it corresponds to the
lower limit of integral $E_{\pi}^{GC}=100~GeV$ in equation (2.5).
The actual value of $E_p$ should be larger than $10^{15}eV$ by one-
to two-magnitude for getting a broad integral range, i.e.,
$E_p=10^{16}\sim 10^{17}eV$. Therefore, we search the GC signal in
AGNs since it is the factory of ultra-high energy protons.

    Conclusions, we find that the GC is one of possible VHE gamma ray sources, which
presents a typical broken power law and caused the hardening
spectrum. The GC occurs at the deepest level of matter, it may
relate to several seemingly unrelated cosmological phenomena. Using
the GC-model we present the following features, which are mentioned
in introduction. (i) The GC-threshold $E_{\pi}^{GC}$ distributes in
a broad region from $GeV$ to $TeV$ energies, which coincides with
the peak range of $\gamma-$ ray distribution; (ii) The cross section
$\sigma_{pp}$ rises linearly with energy rather than increases
logarithmically because the GC-effect contributes all available
kinetic energies to create pions. The channel $pp\rightarrow
\pi^0\rightarrow 2\gamma$ with the GC-effect has a high conversion
efficiency from kinetic energy to $\gamma$-rays; (iii) The broken
power-law is a natural result of the GC-effect in VHE $\gamma$-ray
spectra; (iv) Some of blazars spectra have an extra hard tail at the
TeV range, which can be explained by the contributions of second
GC-source; (v) We present a new dynamic in AGNs, although it has
never been recognized before, but it can be reasoned in QCD; (vi) In
particularly, we expose the GC-effect in the VHE gamma energy
spectra of GRBs. A series of characteristics of the spectra
predicted by the GC model are consistent with the observed results
of GRB 180720B and GRB 190114C. Thus, the GC-effect opens a new
window to understand the anomalous phenomena in cosmic ray energy
spectra.

{\it Acknowledgments:} This work is supported by the National
Natural Science of China (No.11851303). \\\\

\newpage
\begin{table}[htbp]
    \caption{Parameters of  AGN $\gamma-$ray spectra in the GC-model with single source.}\label{tab:1}
    \vskip 0.1cm
    \includegraphics[width=1.05\columnwidth,angle=0]{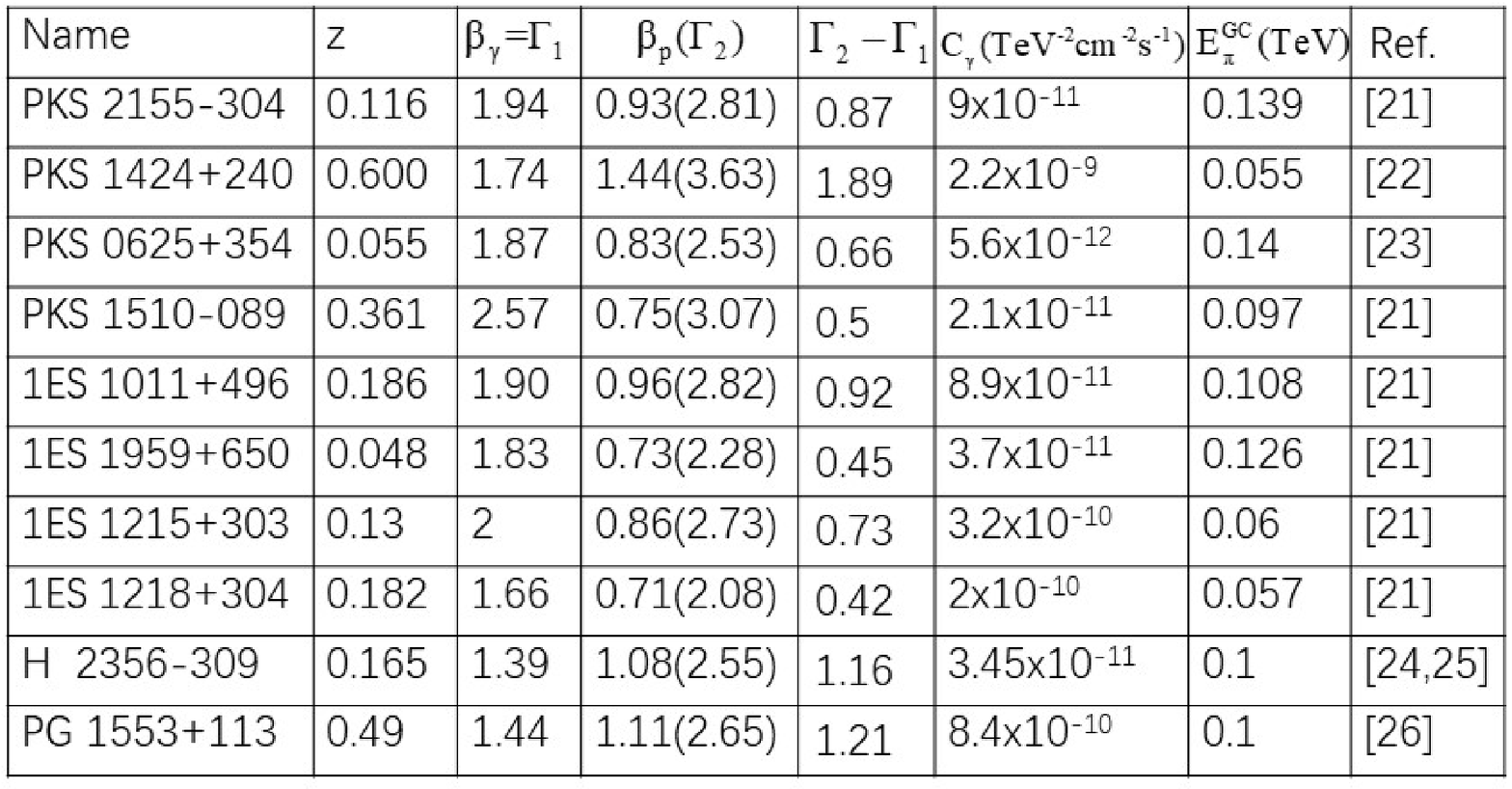}

\end{table}

\begin{table}[htbp]
    \caption{Parameters of  AGN $\gamma-$ray spectra in the GC-model with double sources.}\label{tab:2}
    \vskip 0.04cm
    \includegraphics[width=1.05\columnwidth,angle=0]{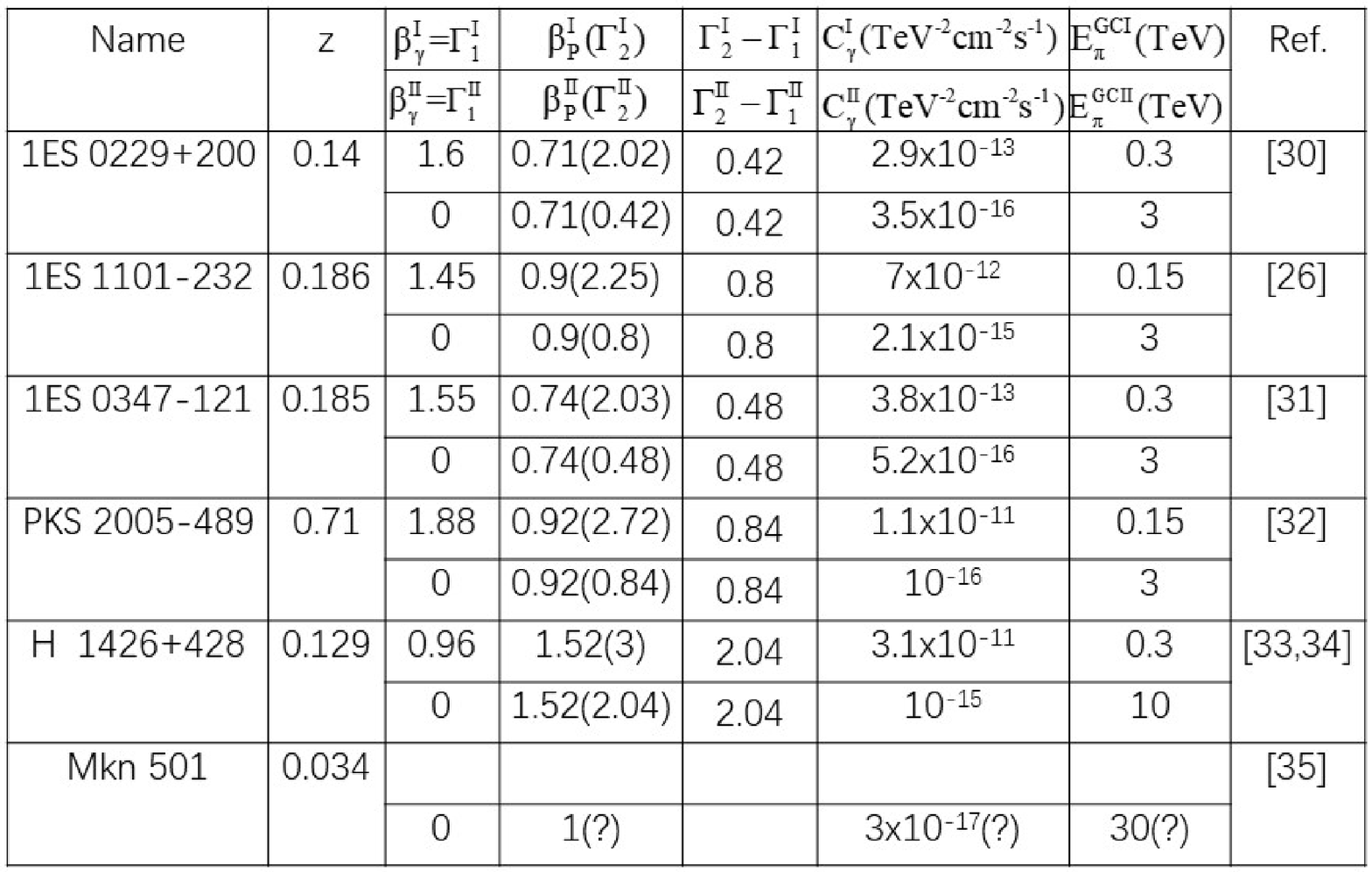}
\end{table}

\begin{figure}[htb]
    \centering {
        \includegraphics[width=0.96\columnwidth,angle=0]{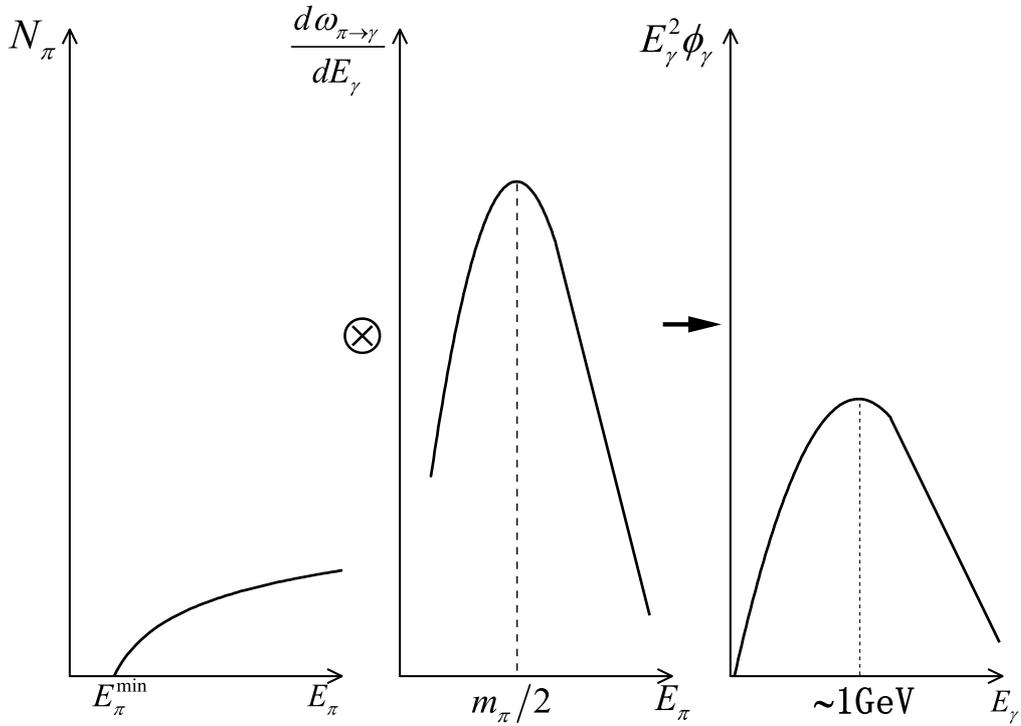}
    } \caption{\label{fig:fig1} Schematic diagrams for $\gamma-$spectra in the hadronic model
    without the GC-effect. Note that all scales are logarithmic.}
\end{figure}

\begin{figure}[htb]
    \centering {
        \includegraphics[width=0.96\columnwidth,angle=0]{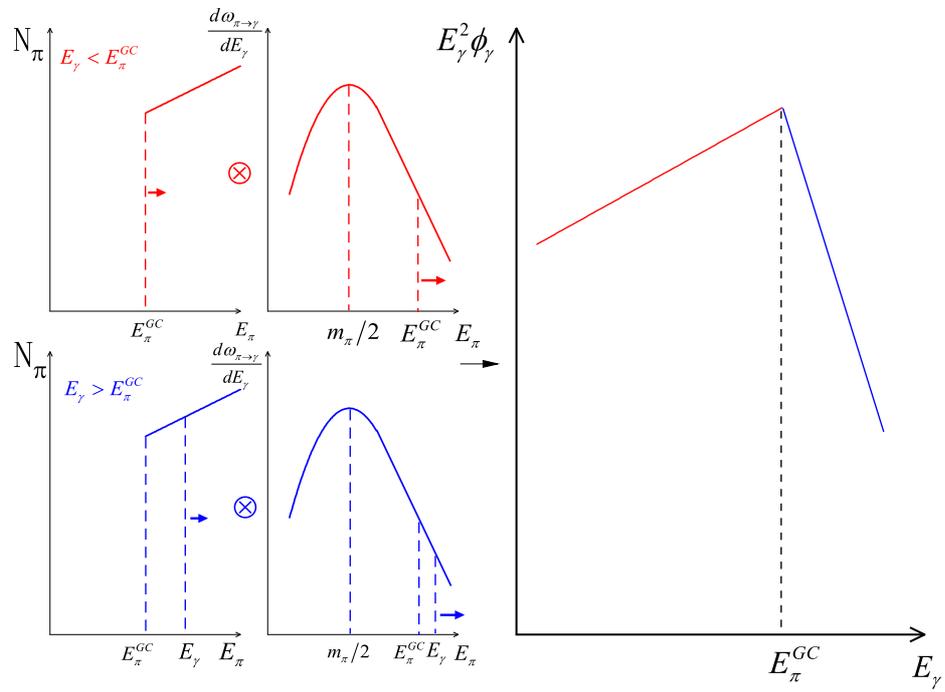}
    } \caption{\label{fig:fig2} Schematic diagrams for $\gamma-$spectra in the hadronic model with
    the GC-effect.}
\end{figure}

\begin{figure}[htb]
    \centering {
        \includegraphics[width=0.96\columnwidth,angle=0]{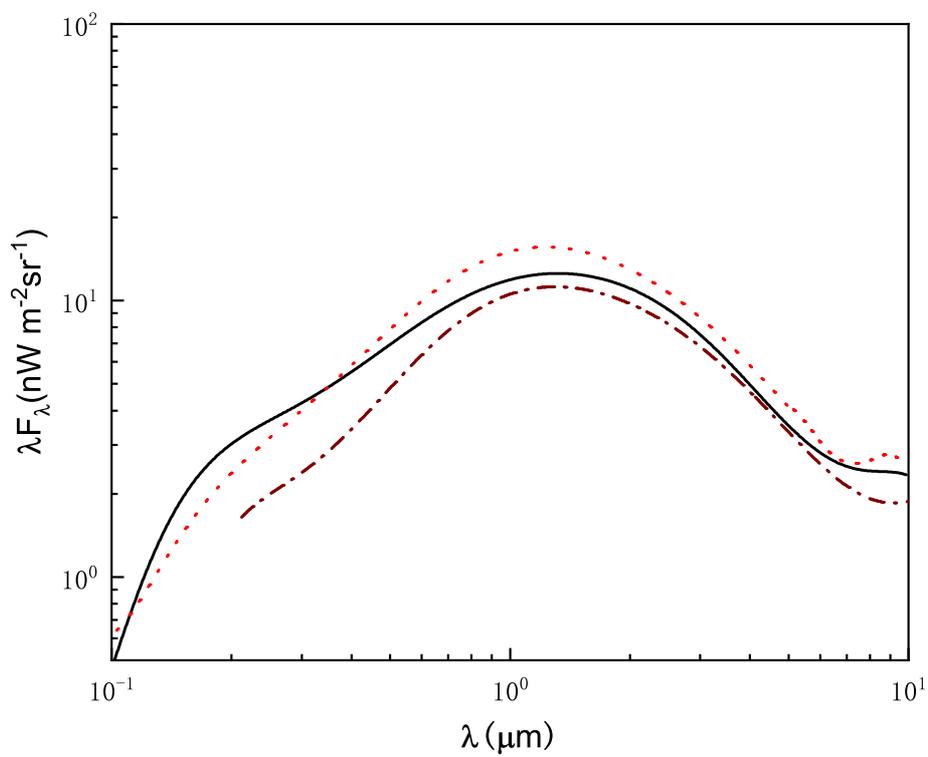}
    } \caption{\label{fig:fig3} Modeled SEDs of the EBL; solid curve is a lower form, which
is proposed by [18] and is used in this work. For comparison, we
give two similar SEDs: dashed and  broken-doted curves are taken
from [19] and [20].}
\end{figure}

\begin{figure}[htb]
    \centering {
        \includegraphics[width=0.96\columnwidth,angle=0]{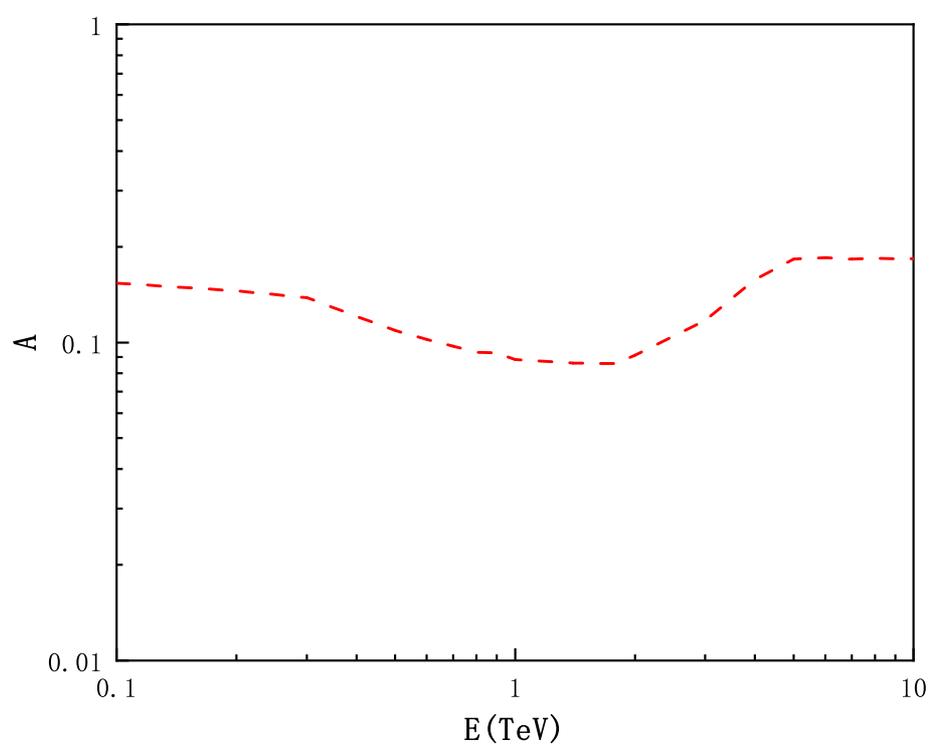}
    } \caption{\label{fig:fig4}  Coefficient A in equation (3.3).}
\end{figure}

\begin{figure}[htb]
    \centering {

        \subfigure{
            \includegraphics[width=2.7in]{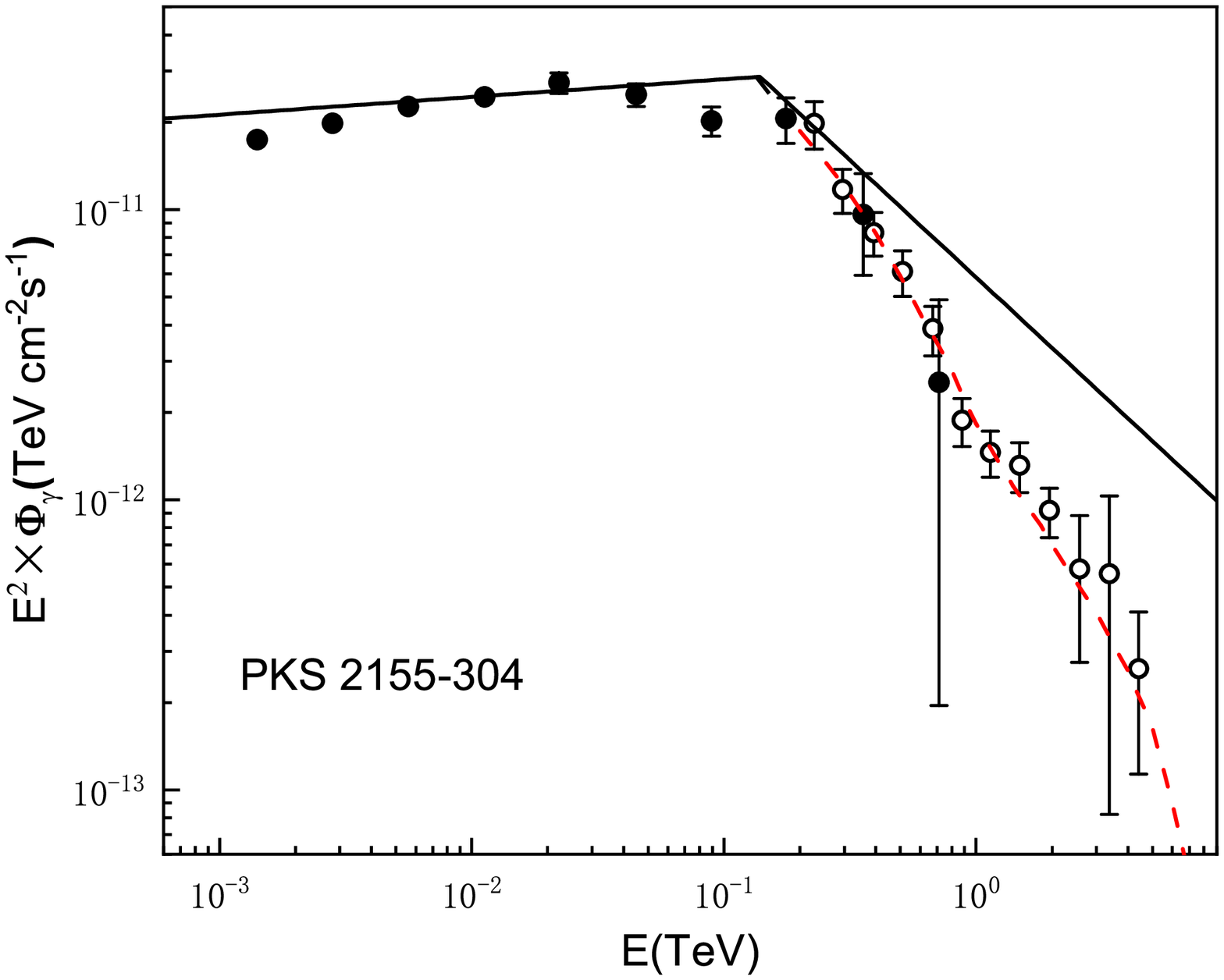} }
        \subfigure{
            \includegraphics[width=2.7in]{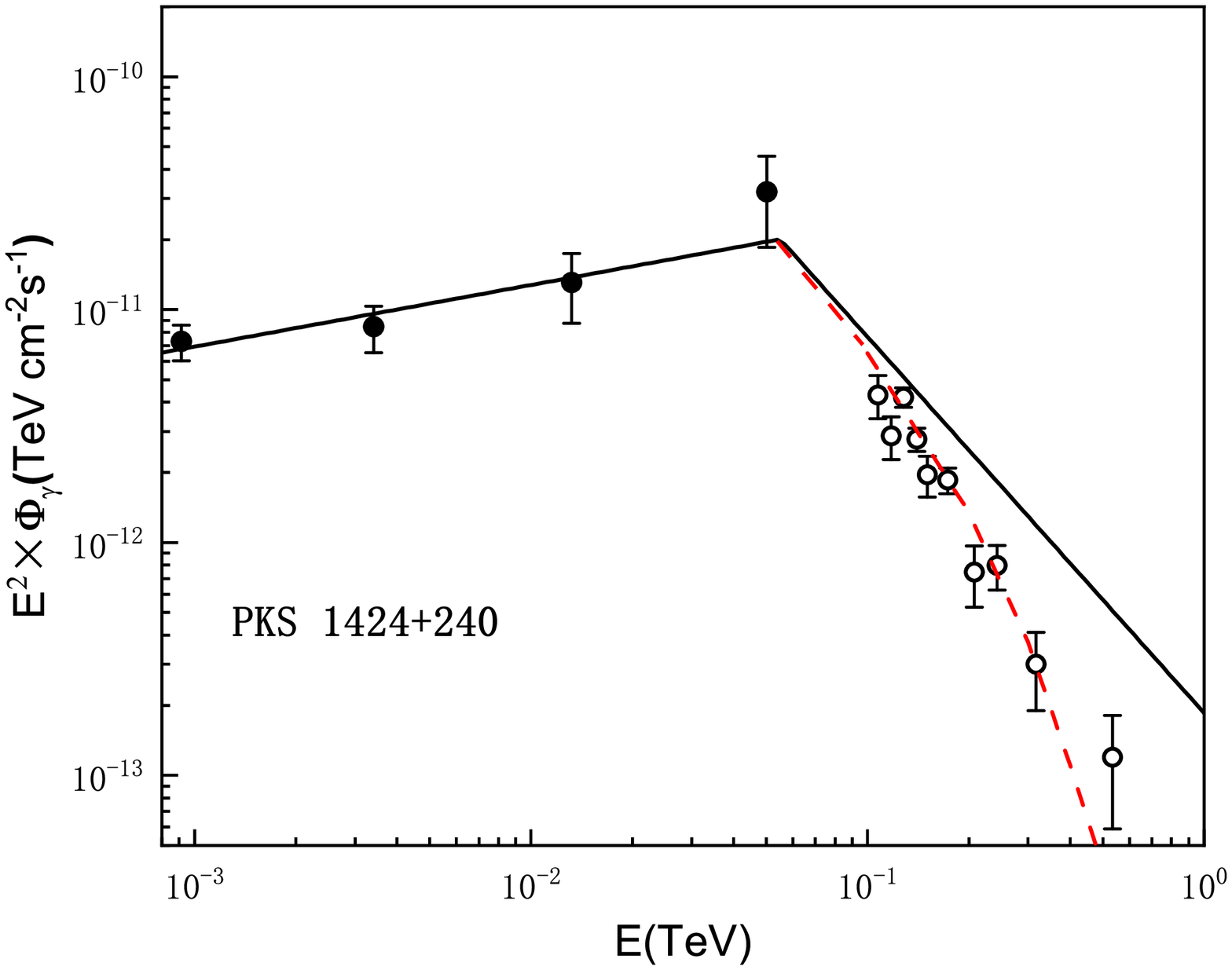}}
        \vskip -0.8cm
        \subfigure{
            \includegraphics[width=2.7in]{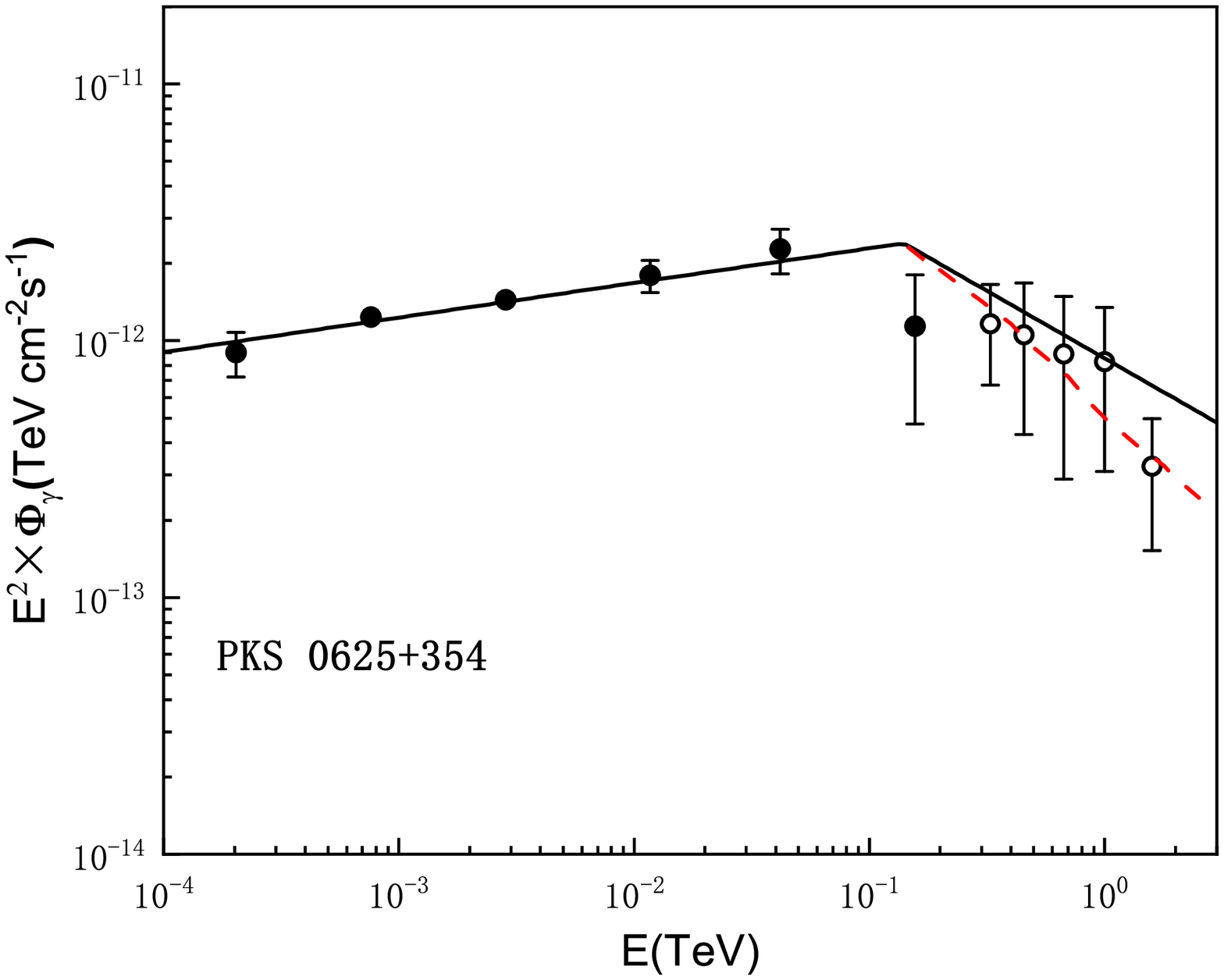} }
        \subfigure{
            \includegraphics[width=2.7in]{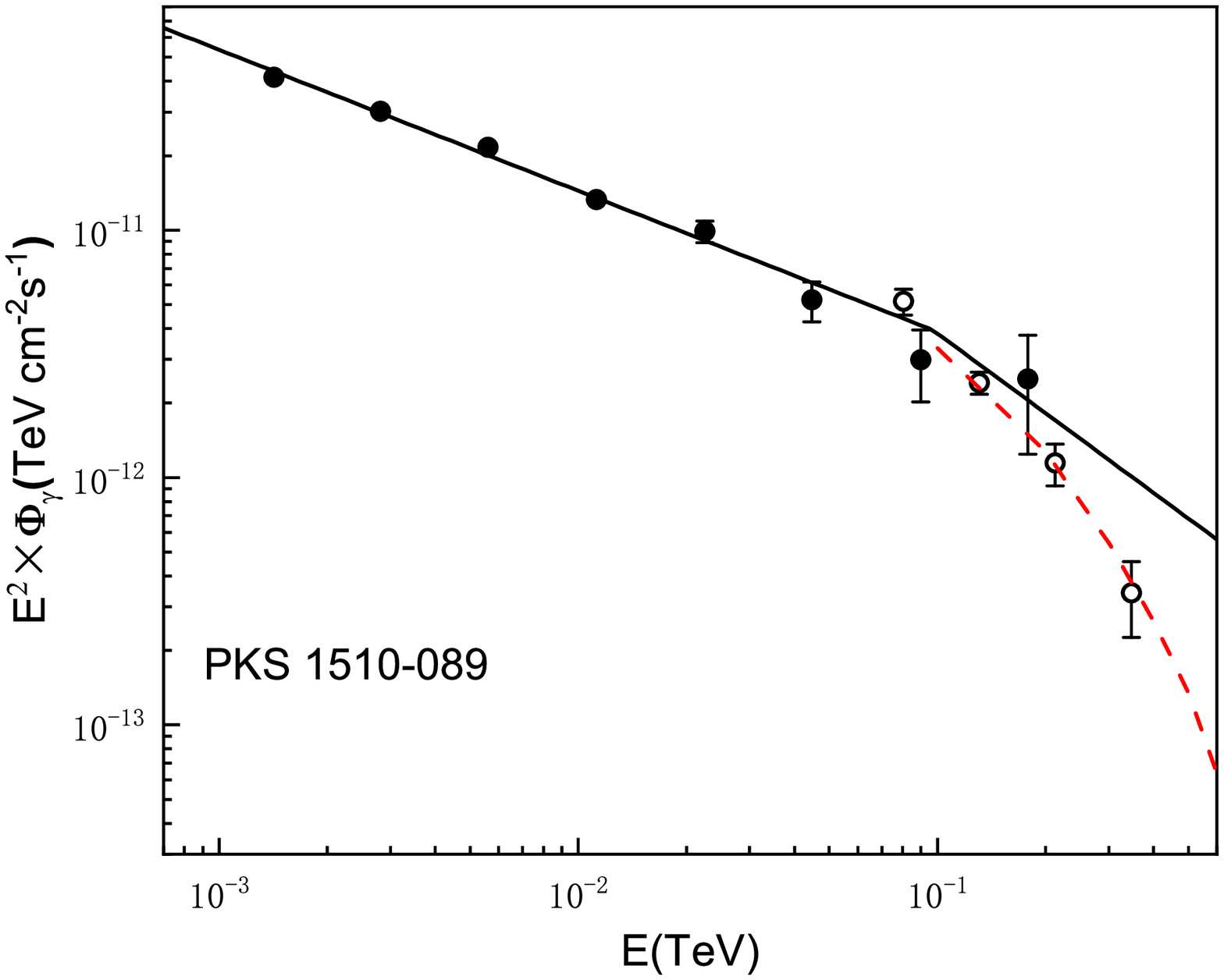}}
        \vskip -0.8cm
        \subfigure{
            \includegraphics[width=2.7in]{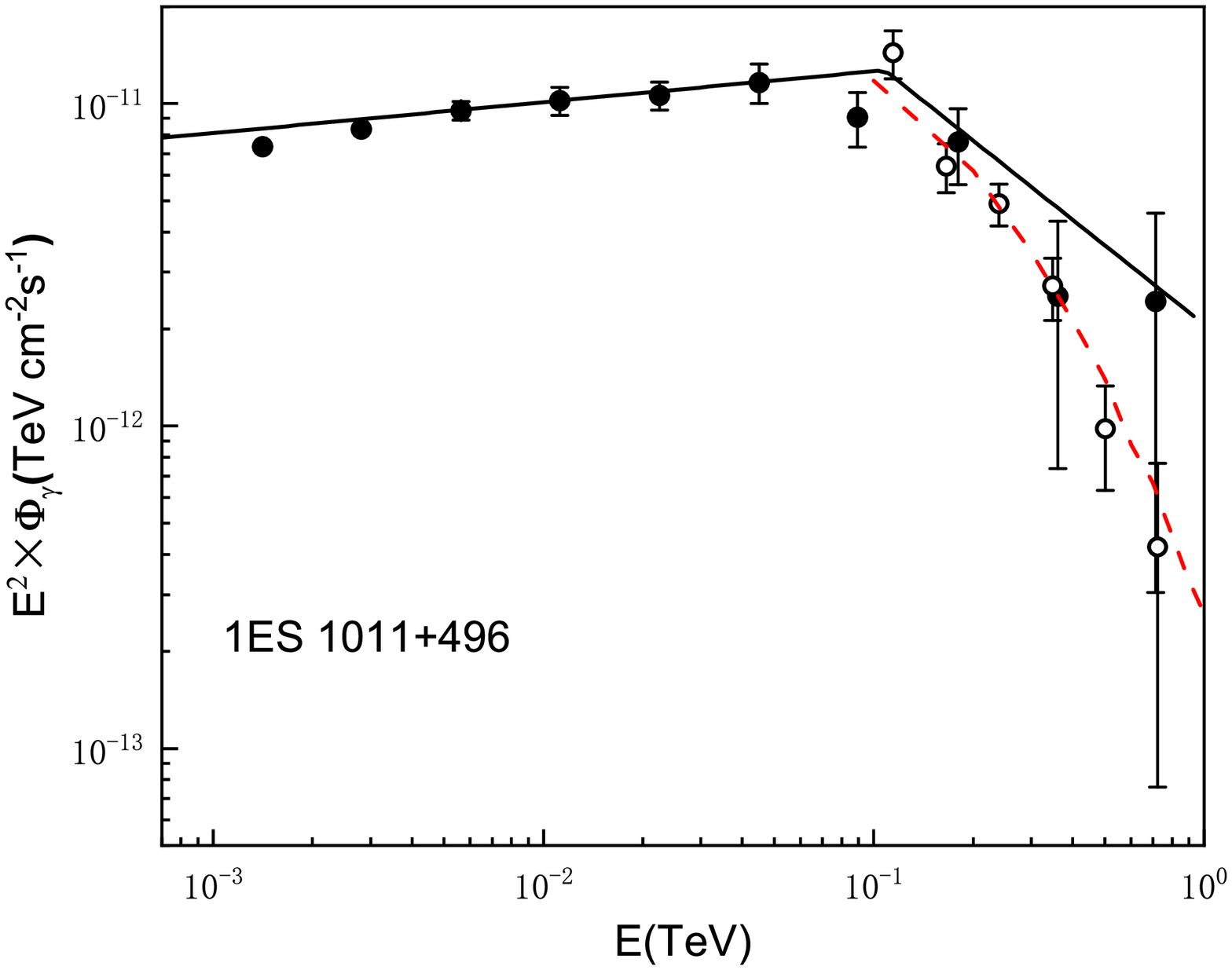} }
        \subfigure{
            \includegraphics[width=2.7in]{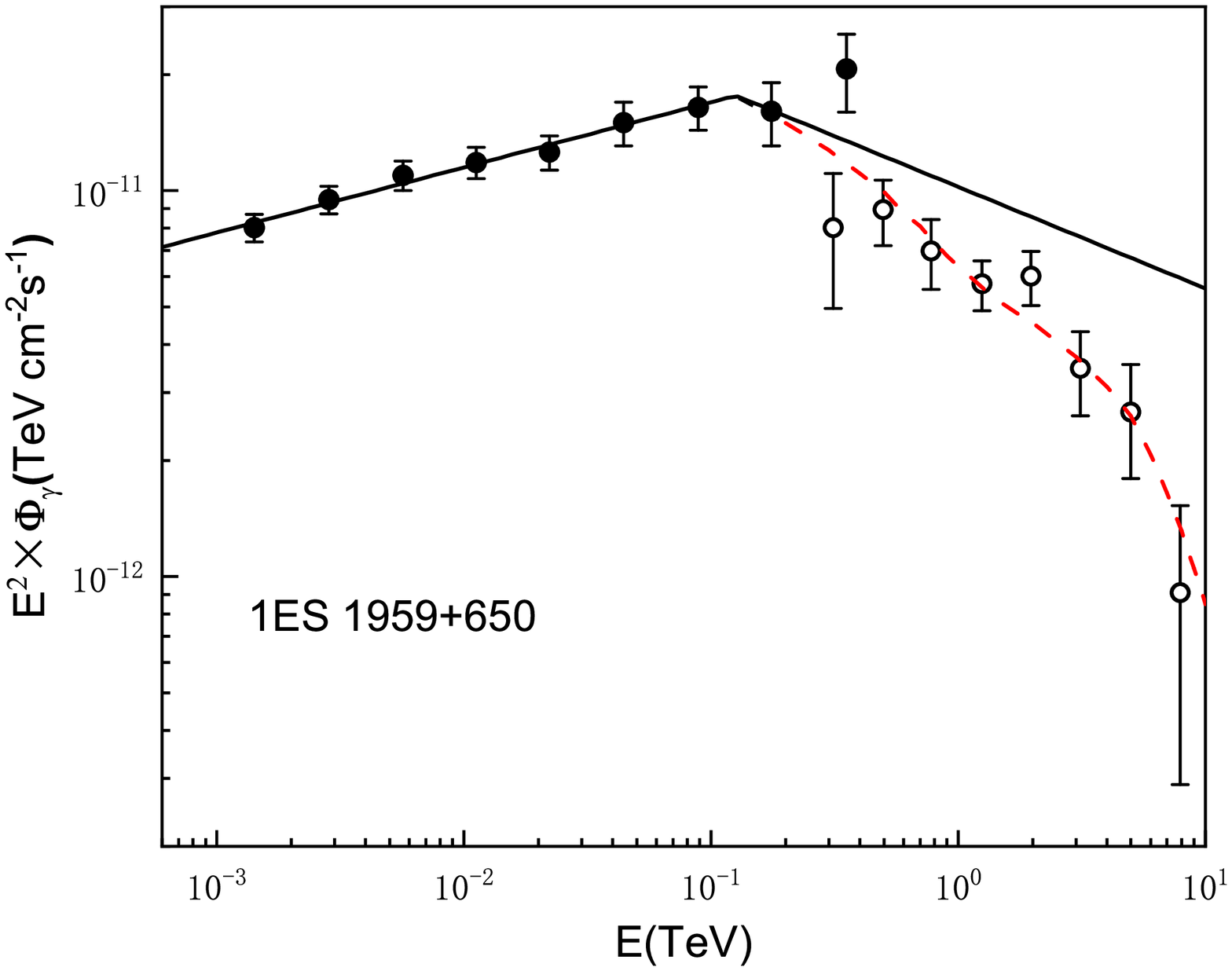}}
        \vskip -0.8cm
        \subfigure{
            \includegraphics[width=2.7in]{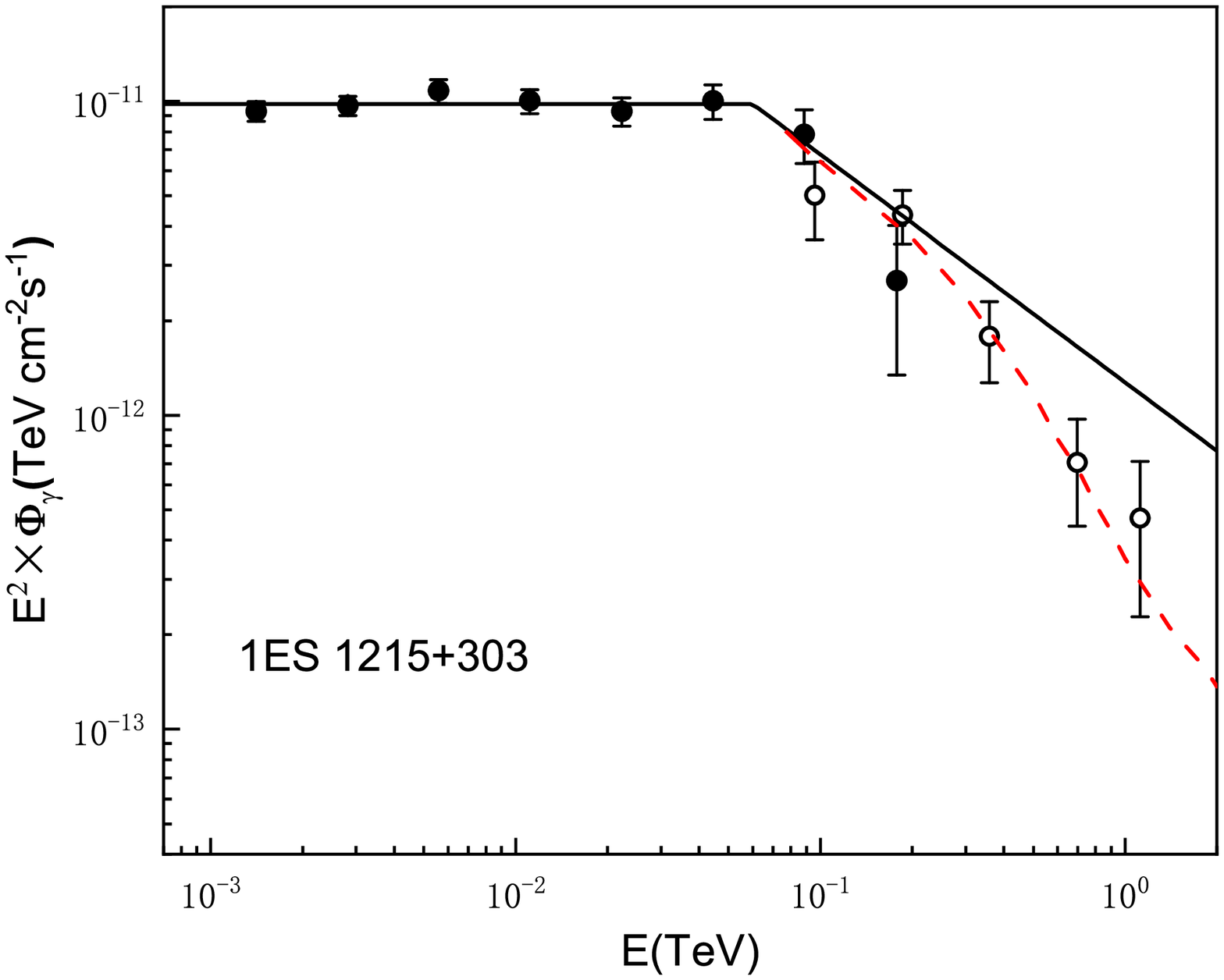} }
        \subfigure{
            \includegraphics[width=2.7in]{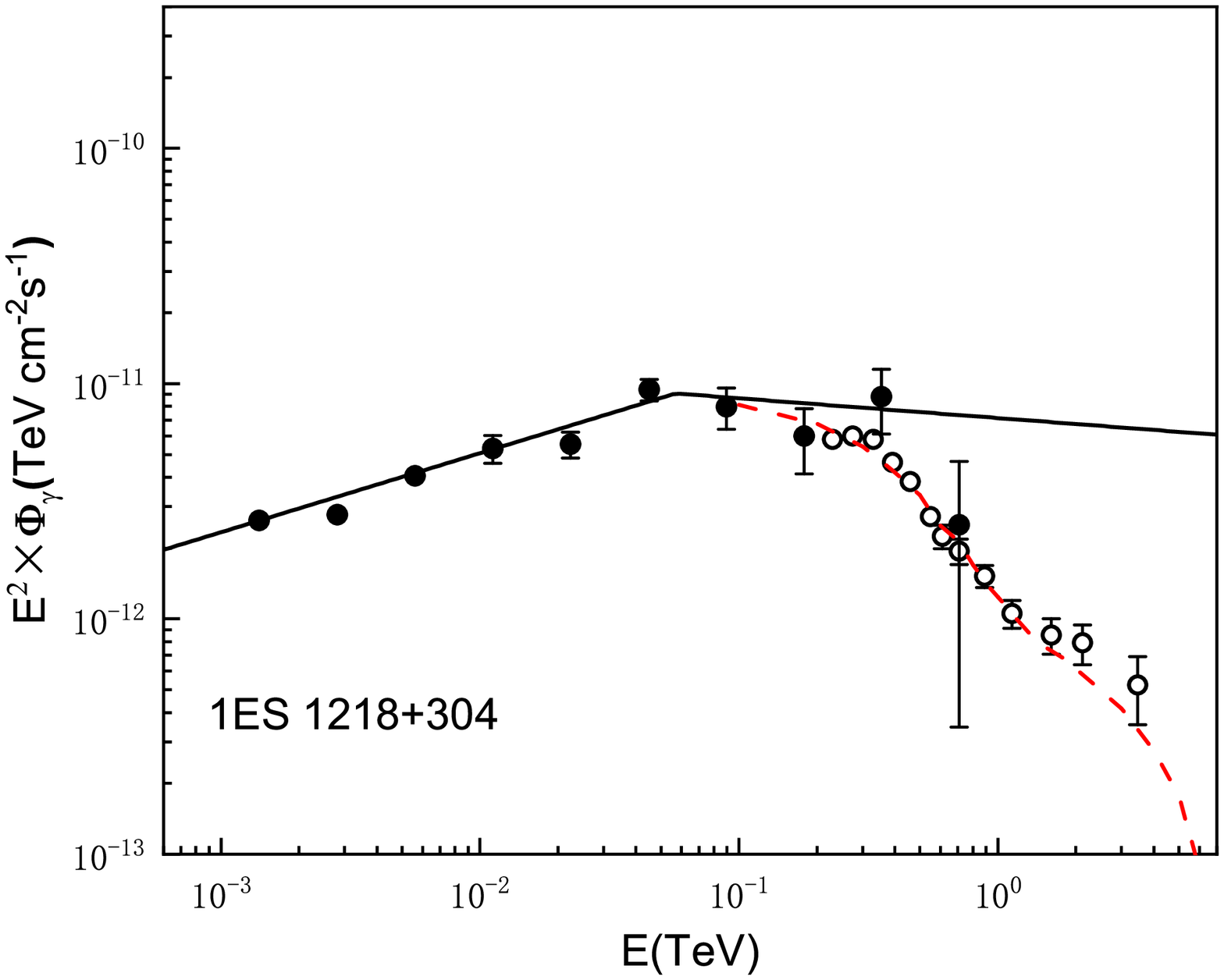}}
        \vskip -0.8cm
        \subfigure{
            \includegraphics[width=2.7in]{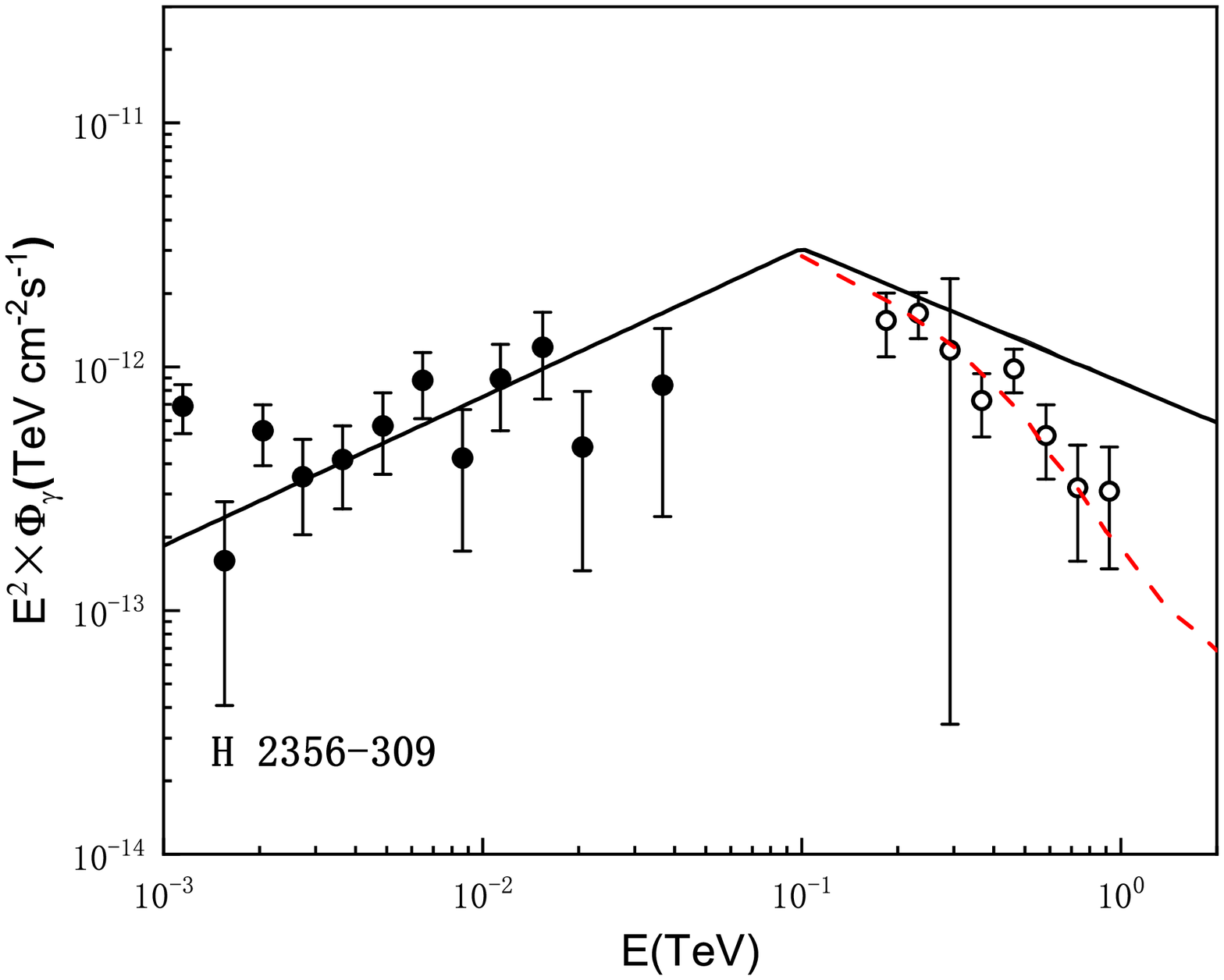} }
        \subfigure{
            \includegraphics[width=2.7in]{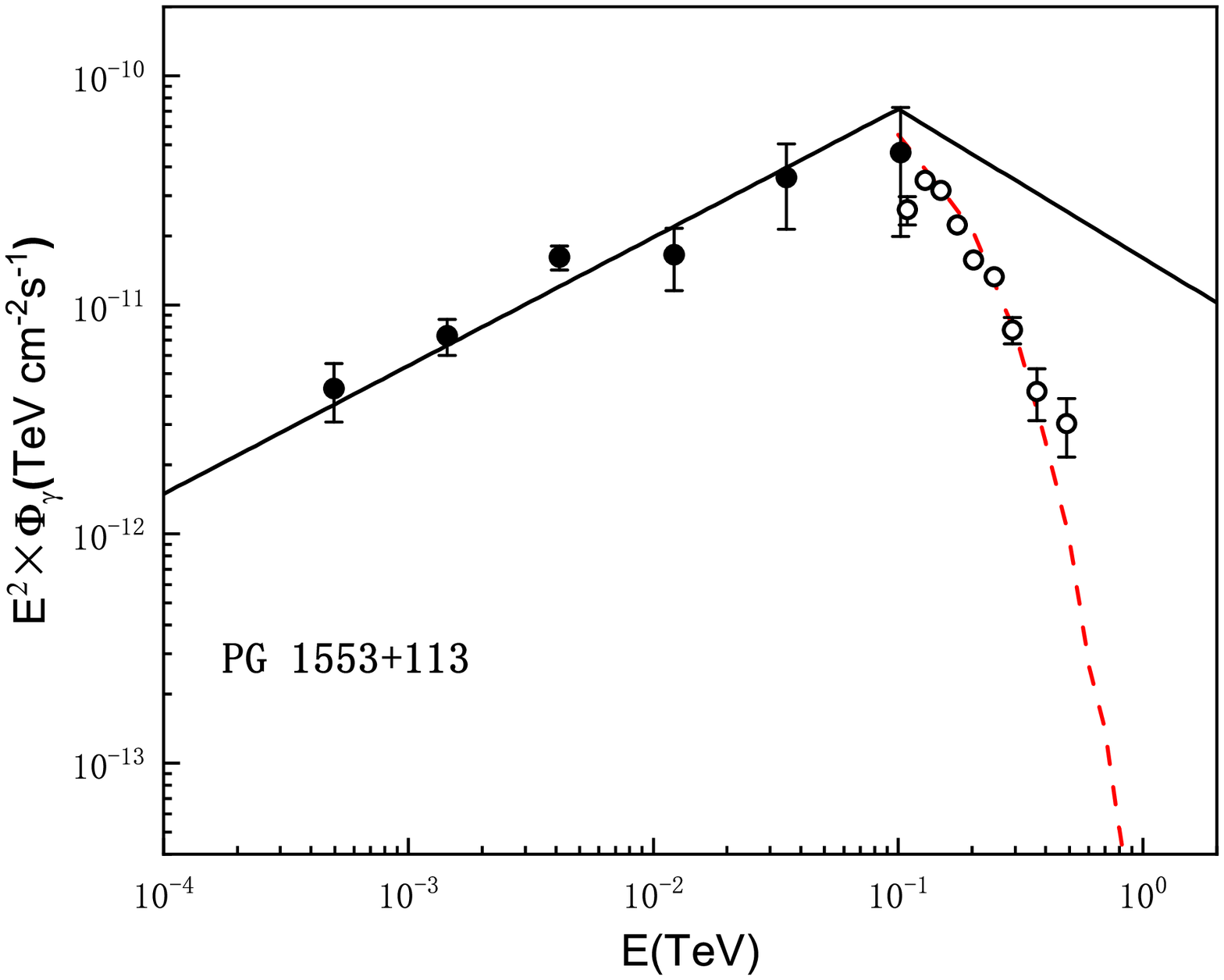}}

    }
    \vskip -0.5cm
    \caption{\label{fig:fig5} SEDs of some TeV blazars. Circles denote measurements;
        the intrinsic spectra of the GC-model are shown in black lines, red dashed
        curves are the observable spectra absorbed with the lower EBL model in figure 3.}

\end{figure}

\begin{figure}[htb]
    \centering {
        \includegraphics[width=0.96\columnwidth,angle=0]{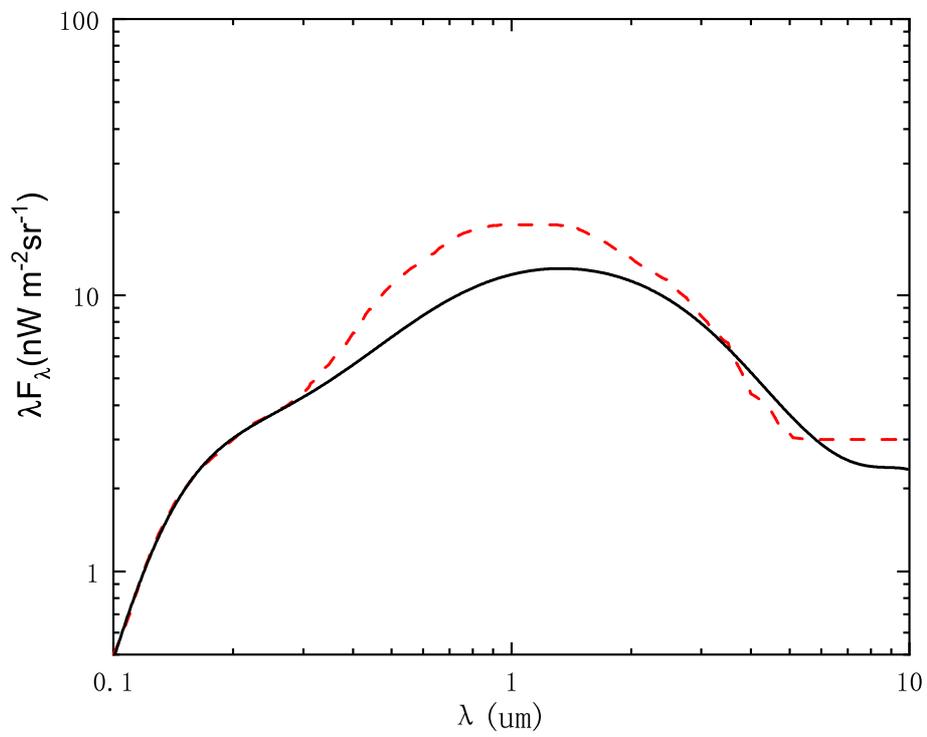}
    } \caption{\label{fig:fig6}  A SED of the EBL (red dashed curve). Black solid curve is
            the lower EBL model in figure 3}
\end{figure}

\begin{figure}[htb]
    \centering {
        \includegraphics[width=0.96\columnwidth,angle=0]{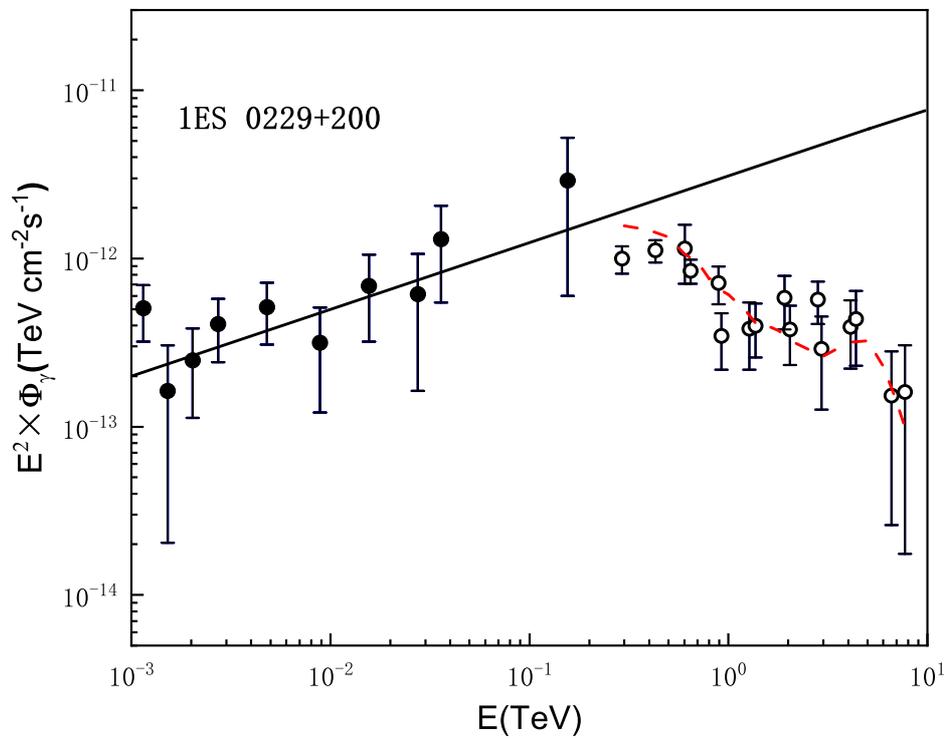}
    } \caption{\label{fig:fig7} SED of 1ES 0229+200, which assumes that
        the break of the spectrum at 0.3 TeV comes entirely from
            a stronger absorption of the EBL in Figure 7. }
\end{figure}

\begin{figure}[htb]
    \centering {

        \subfigure{
            \includegraphics[width=2.7in]{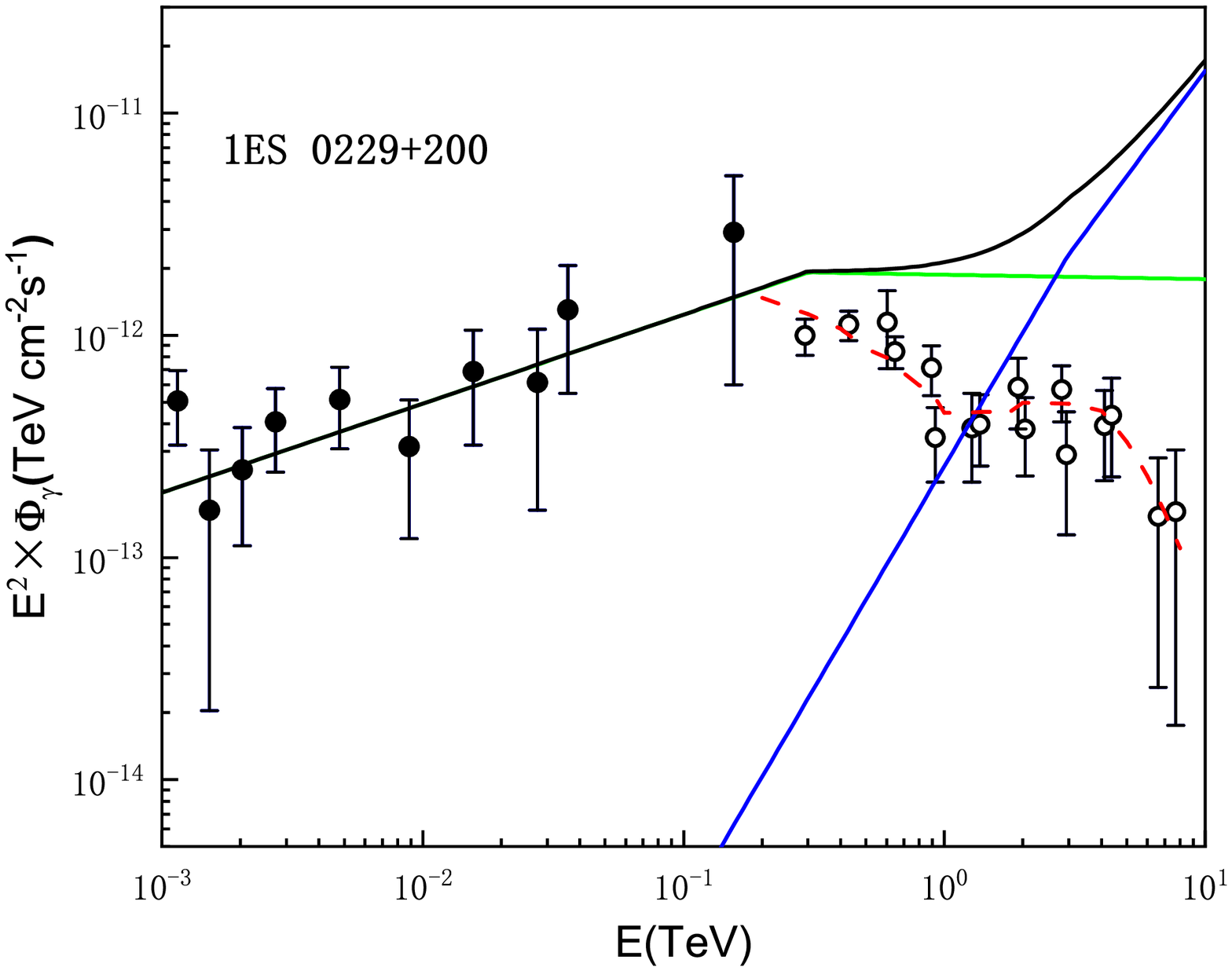} }
        \subfigure{
            \includegraphics[width=2.7in]{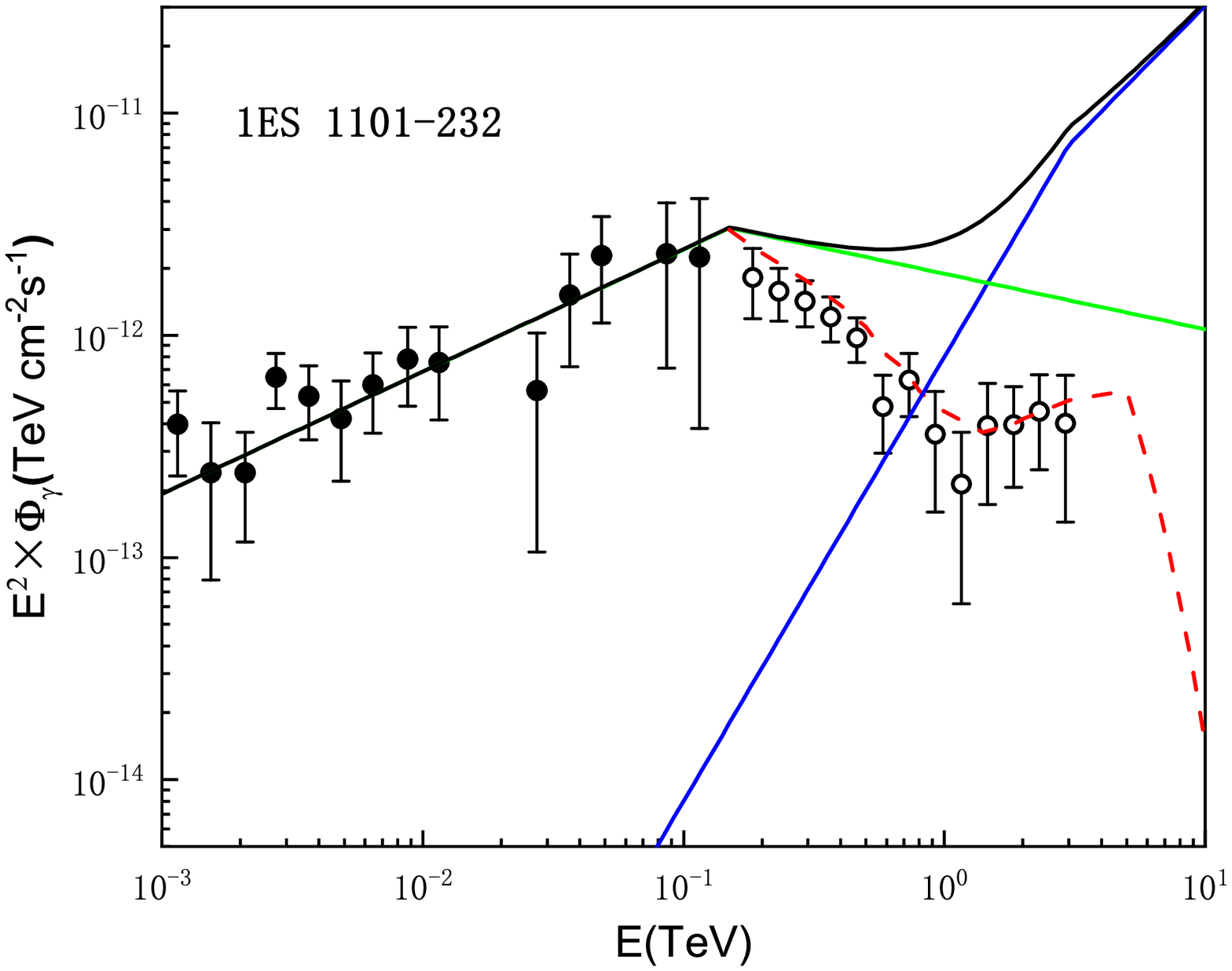}}
        \vskip -0.8cm
        \subfigure{
            \includegraphics[width=2.7in]{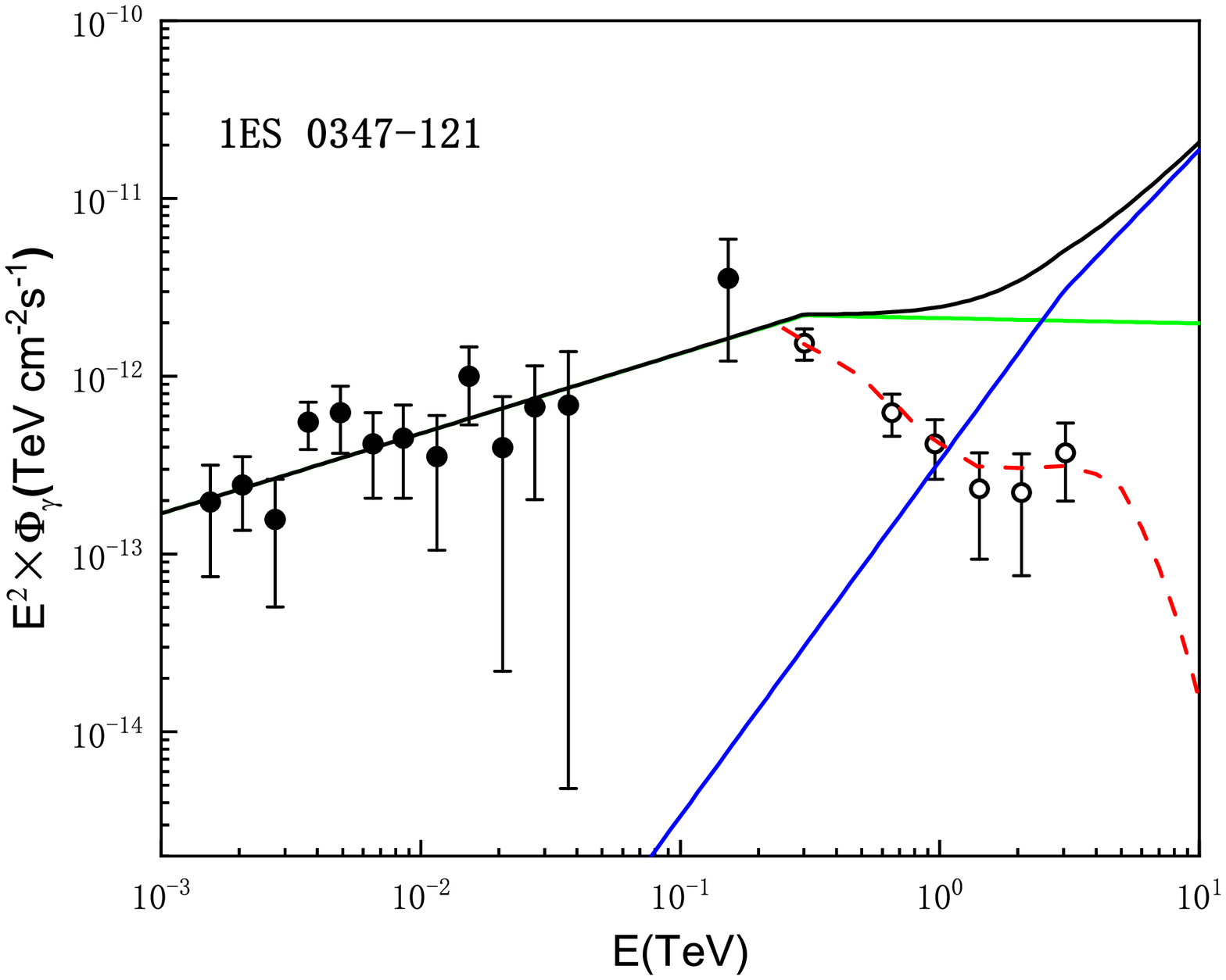} }
        \subfigure{
            \includegraphics[width=2.7in]{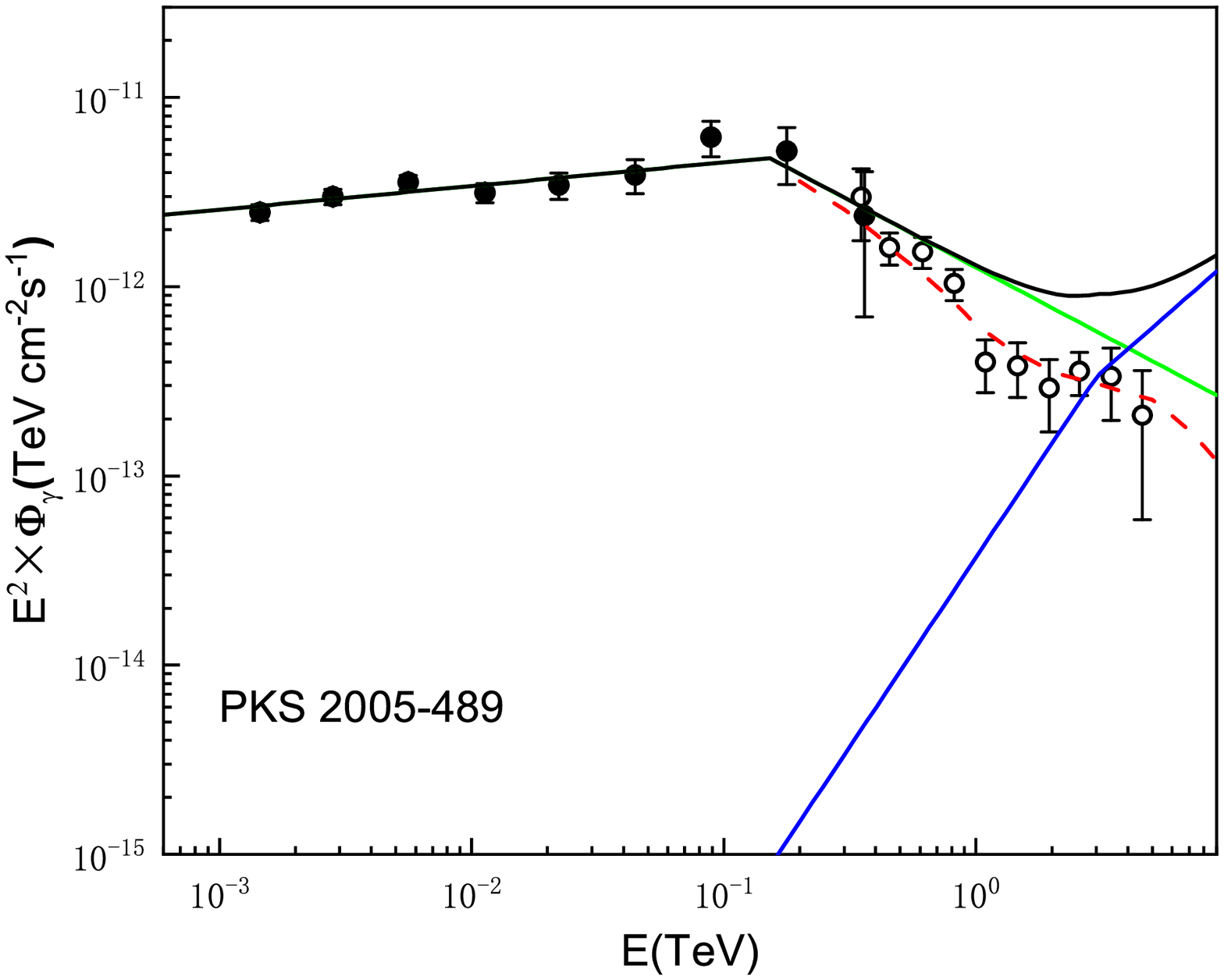}}
        \vskip -0.8cm
        \subfigure{
            \includegraphics[width=2.7in]{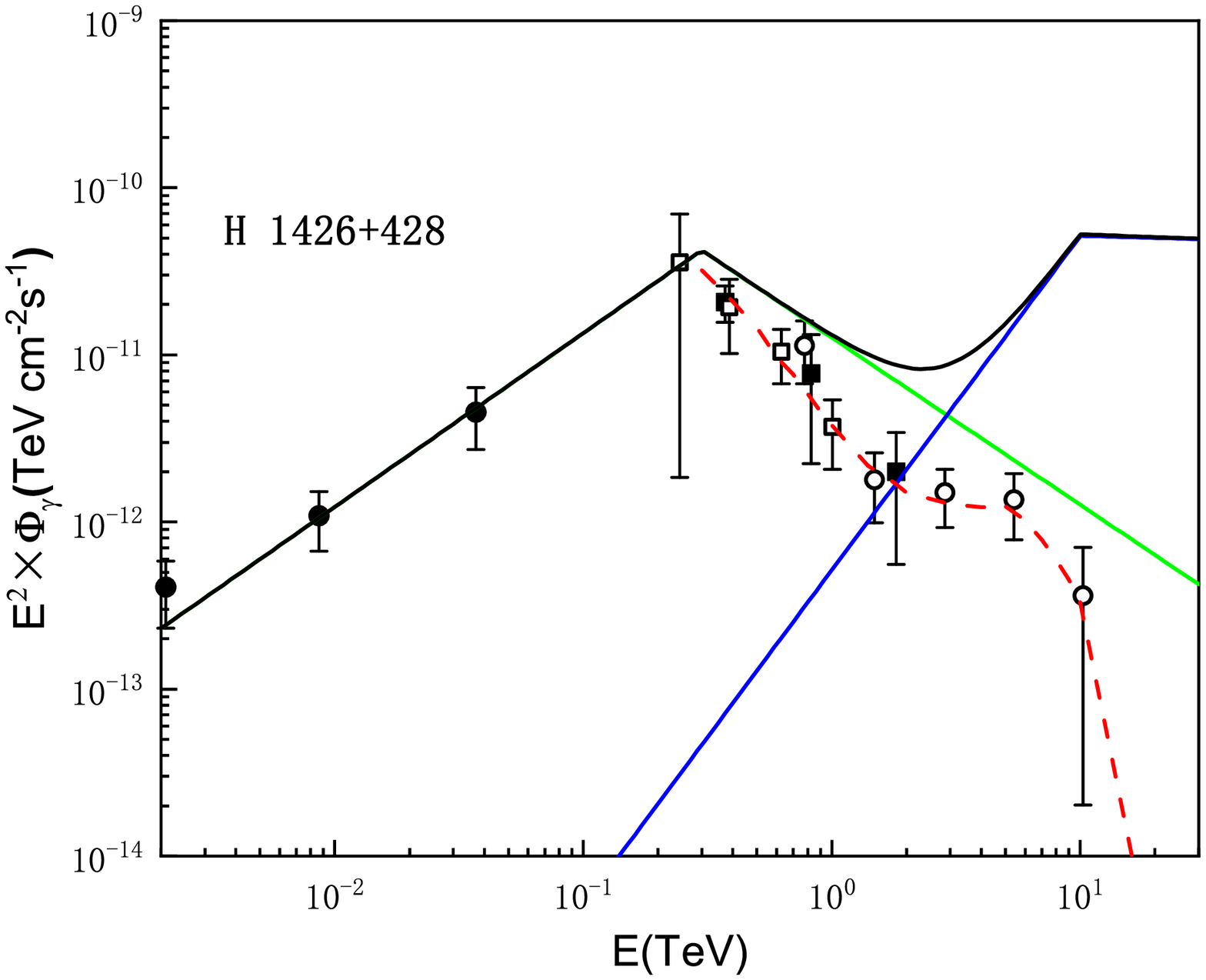} }
        \subfigure{
            \includegraphics[width=2.7in]{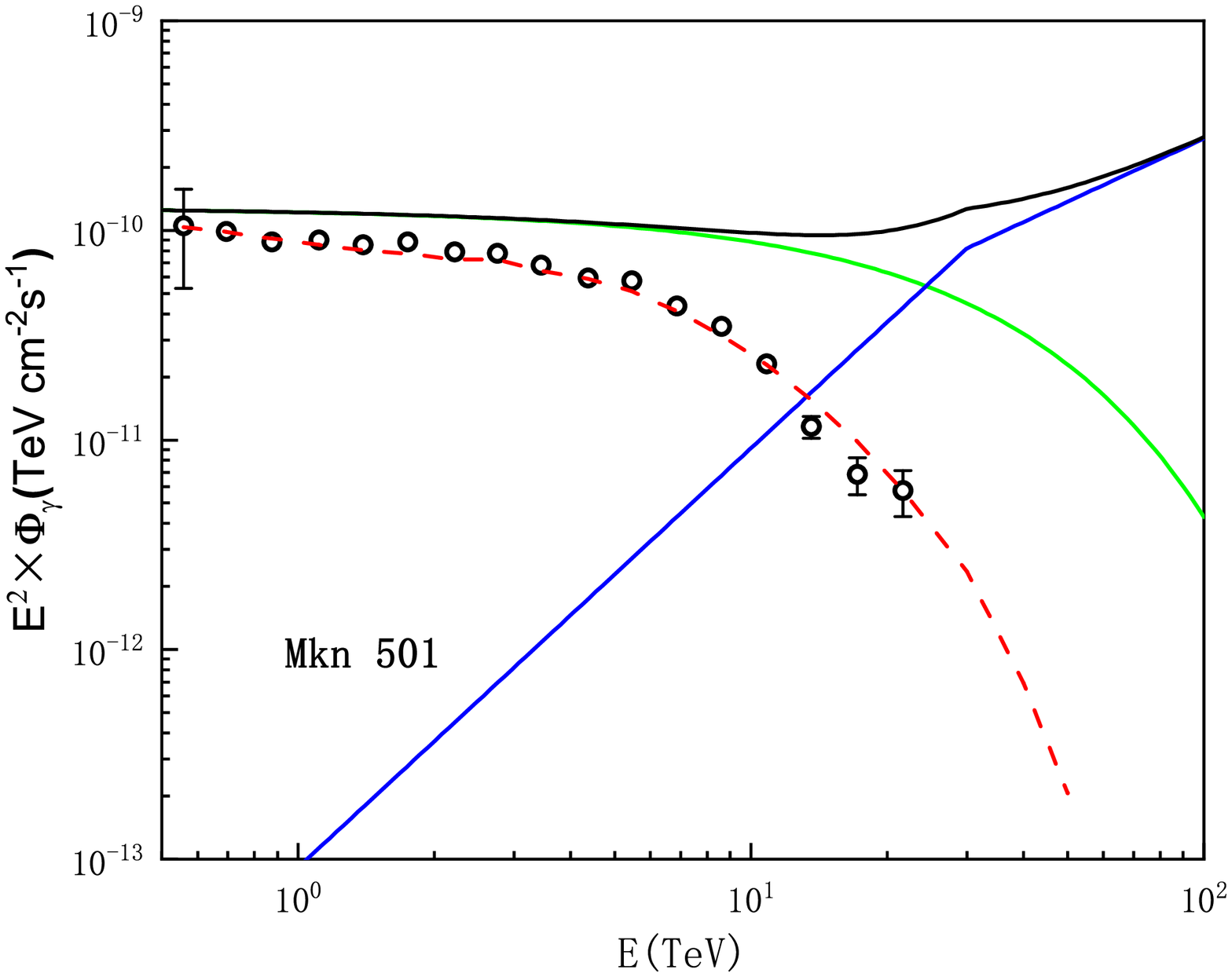}}
    }
    \caption{\label{fig:fig8}Similar to figure 5 but two GC-sources are used,
        where green and blue lines correspond to I- and II-sources, while black curve
        indicates the sum of them. Red dashed curves are the observable
        spectra absorbed with the lower EBL model in figure 3. }

\end{figure}

\begin{figure}[htb]
    \centering {
        \includegraphics[width=0.96\columnwidth,angle=0]{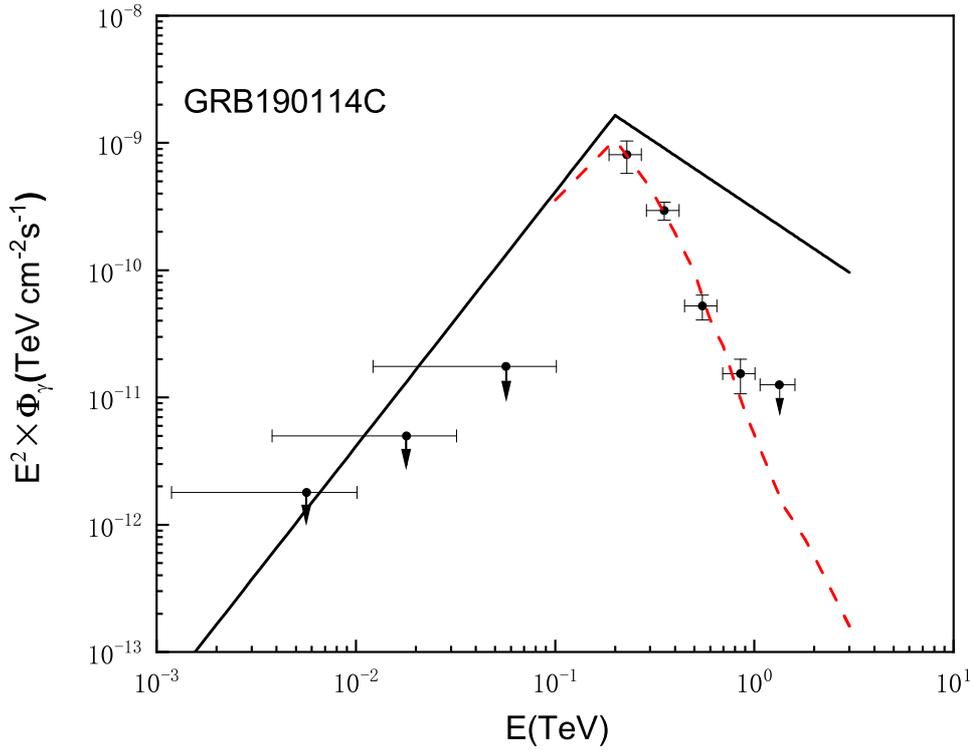}
    } \caption{\label{fig:fig9}
SED of GRB 190114C measured with MAGIC (circles with statistical
uncertainties) [41] together with best fits for the lower EBL model
(dashed curve) and the corresponding intrinsic SEDs (solid curve).
The parameters $\beta_{\gamma}=0, \beta_p=2, E_{\pi}^{GC}=0.2~TeV$
in $\Phi_{\gamma}^{in}$. Note that $\Gamma_2-\Gamma_1=3$. Horizontal
bars with downwards arrows is upper limits derived by [40].}
\end{figure}

\begin{figure}[htb]
    \centering {
        \includegraphics[width=0.96\columnwidth,angle=0]{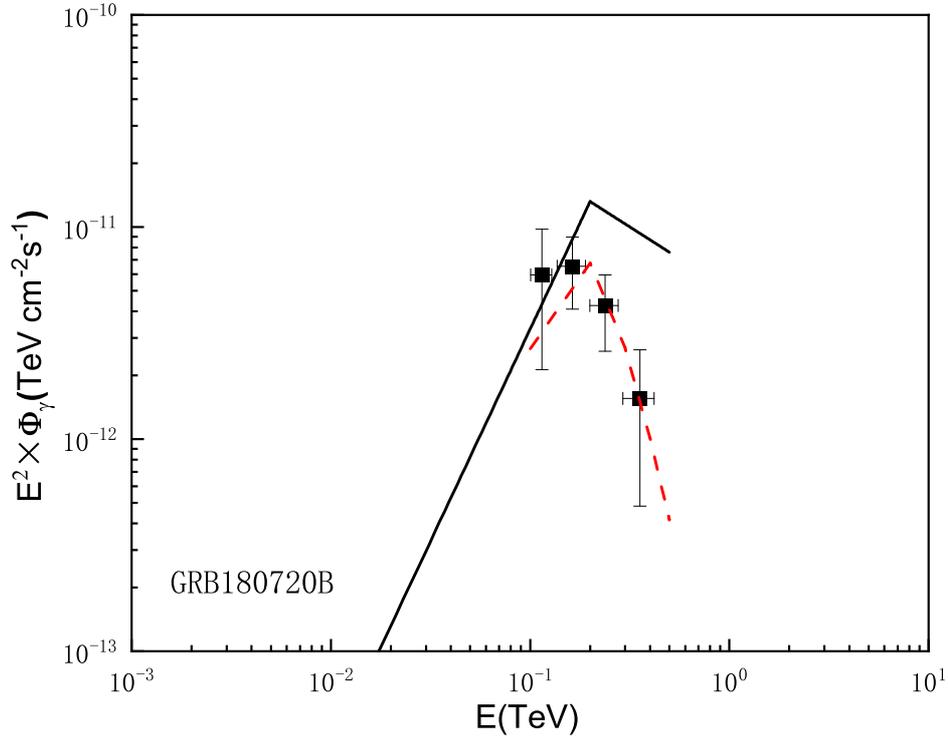}
    } \caption{\label{fig:fig10}SED of GRB 180720B measured with HESS (circles with statistical
uncertainties) [42] together with best fits for the lower EBL model
(dashed curve) and the corresponding intrinsic SEDs (solid curve).
The parameters $\beta_{\gamma}=0, \beta_p=1.8, E_{\pi}^{GC}=0.2~TeV$
in $\Phi_{\gamma}^{in}$. Note that $\Gamma_2-\Gamma_1=2.6$.}
\end{figure}

\begin{figure}[htb]
    \centering {
        \includegraphics[width=0.96\columnwidth,angle=0]{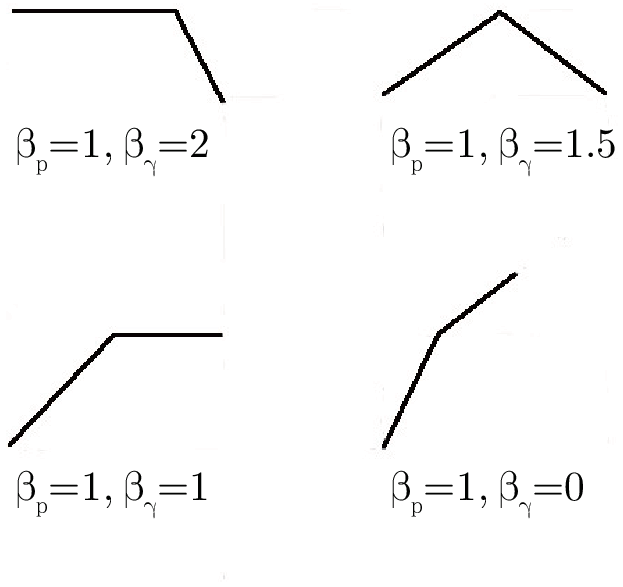}
    } \caption{\label{fig:fig11} Some of spectral forms (multiplied by $E_{\gamma}^2$)
predicted by the GC model. The intrinsic $\gamma$-ray spectra with
these forms can be explained by the GC-model. }
\end{figure}

\end{document}